\documentclass[acmlarge]{acmart}
\usepackage{multirow}
\usepackage{mathtools}
\usepackage[toc,page]{appendix}
\usepackage{amsmath}

\usepackage{lipsum,afterpage,refcount}

\usepackage{float}
\usepackage{caption}

\usepackage{subcaption}
\usepackage{enumitem}
\usepackage{algpseudocode}
\usepackage{algorithm}
\usepackage{xcolor}

\definecolor{myGreen}{HTML}{008000}
\definecolor{myBlue}{HTML}{000090}
\newcommand{\instex}[1]{{#1}}

\usepackage{pgfplots}
 \usepackage{pgfplotstable} 
\pgfplotsset{width=10cm,compat=1.9}

\DeclarePairedDelimiter\ceil{\lceil}{\rceil}

\AtBeginDocument{%
  \providecommand\BibTeX{{%
    \normalfont B\kern-0.5em{\scshape i\kern-0.25em b}\kern-0.8em\TeX}}}

\setcopyright{acmcopyright}
\copyrightyear{2018}
\acmYear{2025}
\acmDOI{10.1145/3712589}

\begin{document}

\title{A Thorough Performance Benchmarking on Lightweight Embedding-based Recommender Systems}

\author{Hung Vinh Tran}
\email{h.v.tran@uq.edu.au}
\affiliation{%
  \institution{The University of Queensland}
  \city{Brisbane}
  \state{Queensland}
  \country{Australia}
  \postcode{4072}
}
\author{Tong Chen}
\email{tong.chen@uq.edu.au}
\affiliation{%
  \institution{The University of Queensland}
  \city{Brisbane}
  \state{Queensland}
  \country{Australia}
  \postcode{4072}
}
\author{Nguyen Quoc Viet Hung}
\email{henry.nguyen@griffith.edu.au}
\affiliation{%
  \institution{Griffith University}
  \city{Gold Coast}
  \state{Queensland}
  \country{Australia}
  \postcode{4222}
}
\author{Zi Huang}
\email{helen.huang@uq.edu.au}
\affiliation{%
  \institution{The University of Queensland}
  \city{Brisbane}
  \state{Queensland}
  \country{Australia}
  \postcode{4072}
}
\author{Lizhen Cui}
\authornote{Corresponding authors.}
\email{clz@sdu.edu.cn}
\affiliation{%
  \institution{Shandong University}
  \city{Jinan}
  \state{Shandong}
  \country{China}
  \postcode{250000}
}
\author{Hongzhi Yin}
\authornotemark[1]
\email{h.yin1@uq.edu.au}
\affiliation{%
  \institution{The University of Queensland}
  \city{Brisbane}
  \state{Queensland}
  \country{Australia}
  \postcode{4072}
}

\renewcommand{\shortauthors}{Tran, et al.}

\begin{abstract}
Since the creation of the Web, recommender systems (RSs) have been an indispensable personalization mechanism in information filtering. 
Most state-of-the-art RSs primarily depend on categorical features such as user and item IDs, and use embedding vectors to encode their information for accurate recommendations, resulting in an excessively large embedding table owing to the immense feature corpus. 
To prevent the heavily parameterized embedding table from harming RSs' scalability, both academia and industry have seen increasing efforts compressing RS embeddings, and this trend is further amplified by the recent uptake in edge computing for online services.
However, despite the prosperity of existing lightweight embedding-based RSs (LERSs), a strong diversity is seen in the evaluation protocols adopted across publications, resulting in obstacles when relating the reported performance of those LERSs to their real-world usability. 
On the other hand, among the two fundamental recommendation tasks, namely traditional collaborative filtering and content-based recommendation, despite their common goal of achieving lightweight embeddings, the outgoing LERSs are designed and evaluated with a straightforward ``either-or'' choice between the two tasks. 
Consequently, the lack of discussions on a method's cross-task transferability will likely hinder the development of unified, more scalable solutions for production environments. 
Motivated by these unresolved issues, this study aims to systematically investigate existing LERSs' performance, efficiency, and cross-task transferability via a thorough benchmarking process.
To create a generic, task-independent baseline, we propose an efficient embedding compression approach based on magnitude pruning, which is proven to be an easy-to-deploy yet highly competitive baseline that outperforms various complex LERSs. 
Our study reveals the distinct performance of different LERSs across the two recommendation tasks, shedding light on their effectiveness and generalizability under different settings.
Furthermore, to account for edge-based recommendation -- an increasingly popular use case of LERSs, we have also deployed and tested all LERSs on a Raspberry Pi 4, where their efficiency bottleneck is exposed compared with GPU-based deployment. 
Finally, we conclude this paper with critical summaries on the performance comparison, suggestions on model selection based on task objectives, and underexplored challenges around the applicability of existing LERSs for future research. 
To encourage and support future LERS research, we publish all source codes and data, checkpoints, and documentation at \href{https://github.com/chenxing1999/recsys-benchmark}{https://github.com/chenxing1999/recsys-benchmark}.
\end{abstract}

\ccsdesc[500]{Information systems~Recommender systems}

\keywords{Benchmarking, On-device Recommendation, Lightweight Recommender Systems.}




\maketitle

    


\section{Introduction}\label{sec:intro}

    
Nowadays, recommender systems (RSs) play a crucial role in assisting users to identify relevant information in their daily lives. A study \cite{memcom2020} shows that RSs contributed substantially across various platforms, including 35\% of Amazon's revenue, 23.7\% of BestBuy's growth, as well as up to 75\% and 60\% views on Netflix and Youtube, respectively.
Since most modern RSs predominantly rely on categorical features, such as user occupations and movie tags in content-based recommendation \cite{deepfm2017}, a crucial way to ensure model expressiveness is to represent each categorical feature as a unique embedding. The embeddings are generally defined as fixed-length vectors for the ease of downstream computations (e.g., dot product for modeling feature interactions or user-item affinity), and are hosted by the RS model in a dense matrix form, commonly referred to as the embedding table.

Given the huge amount and diversity of real-world categorical features, embedding tables consequently dominate the parameter consumption in many modern RSs \cite{zhang2023experimental}. For example, LightGCN \cite{lightgcn2020}, a graph-based collaborative filtering model, spends all parameters on its embedding table, while DeepFM \cite{deepfm2017}, a representative content-based recommender, relies on an embedding table that consumes more than 80\% of its parameters for the Criteo benchmark dataset \cite{Criteo}.
Another concrete example is the RS deployed by Meta, which consumes 12T parameters and can demand up to 96TB of memory and multiple GPUs to train \cite{metarecsys2022}. The bulkiness of an RS's embedding table has consequently triggered the recent proliferation of lightweight embedding-based recommender systems (LERSs), which have seen popularity in both research \cite{long2023model, han2021deeprec} and industry deployments \cite{memcom2020, lam2023gpu}. On the one hand, the reduced parameterization in the embedding table provides a direct cure to the scalability bottleneck of large-scale RSs, where some work has reported a 10$\times$ parameter reduction with negligible performance compromise \cite{optembed2022,ttrec2021}. On the other hand, this line of research also co-evolves with the ongoing trend of decentralization in the RS service architecture, which provides various benefits such as low latency, better privacy, and reduced server hosting costs. Federated \cite{ammad2019federated, chai2020secure}, on-device \cite{rule2021, qin2023learning,yin2024device}, and IoT-enhanced \cite{iotrs-survey-2020} recommendation paradigms all align with this decentralization trend, while they unanimously put a higher demand for nimble RS solutions that can operate on less resourceful edge devices (e.g., a smartphone) -- hence the rise of LERSs.

\begin{figure}
    \centering
    \subcaptionbox{Original}
    {
        \vspace{-.2cm}
        \includegraphics[width=0.2073\textwidth]{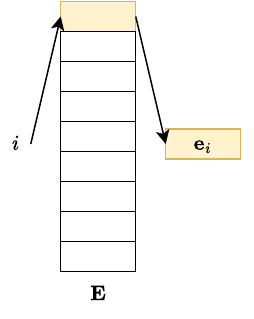}
    }
    \hfill
    \subcaptionbox{Compositional}
    {
        \vspace{-.2cm}
        \includegraphics[width=0.2461\textwidth]{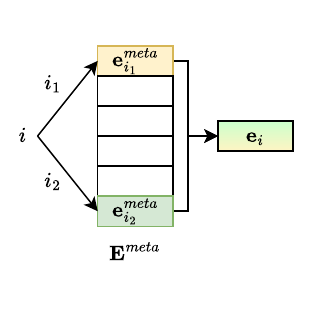}
    }
    \hfill
    \subcaptionbox{Pruning}
    {
        \vspace{-.2cm}
        \includegraphics[width=0.1943\textwidth]{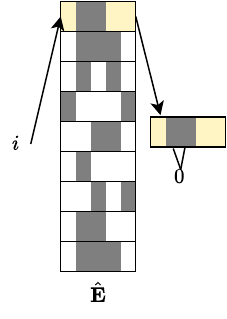}
    }
    \hfill
    \subcaptionbox{NAS-based}
    {
        \vspace{-.2cm}
        \includegraphics[width=0.2721\textwidth]{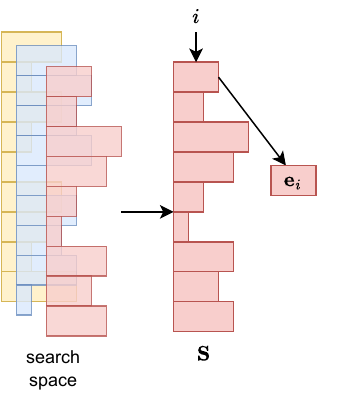}
    }

    
    \caption{
    \textbf{Illustration for main archetypes of LERSs embedding. }
    \instex{\textbf{(a)} The original embedding table uses a dedicated vector $\mathbf{e}_i$ from table $\mathbf{E}$ to represent feature $i$.
    \textbf{(b)} The compositional embedding approach employs a smaller embedding table $\mathbf{E}^{meta}$ and a hash function that hashes $i$ into $(i_1, i_2)$. The final embedding $\mathbf{e}_i$ is constructed by combining $\mathbf{e}^{meta}_{i_1}$ and $\mathbf{e}^{meta}_{i_2}$.
    \textbf{(c)} Pruning reduces the size of the original embedding table by zeroing out parts of it (shown in gray), resulting in a sparse embedding table $\mathbf{\hat{E}}$ with a smaller storage footprint.
    \textbf{(d)} Methods based on neural architecture search (NAS) commonly search for an optimal embedding dimension configuration $S$ for the embedding table from a given search space, optimizing the trade-off between memory cost and model performance.} 
    }
    \label{fig:related-works}
\end{figure}

Despite the promising blueprint, providing accurate recommendations is challenging under limited memory and computing budgets \cite{ondevicesurvey2021}. For LERSs, researchers proposed diverse approaches to compress the embedding table of recommender systems, as illustrated in Figure \ref{fig:related-works}.(a). We hereby summarize the three main categories of methods for achieving lightweight yet expressive embeddings for RSs, which are also depicted in Figure \ref{fig:related-works}.(b)-(d):
\begin{enumerate}
    \item \textbf{Compositional embedding} (Figure \ref{fig:related-works}.(b)) leverages one or multiple smaller embedding tables (also called meta-embedding tables \cite{cerp2023}), where each feature is represented with a unique combination of meta-embeddings. Various operators (e.g., sum, element-wise multiply, concatenate) can be used to transform a collection of meta-embeddings into a single, unique embedding for each feature \cite{qr2019, ttrec2021}. 
    \item \textbf{Embedding pruning} (Figure \ref{fig:related-works}.(c)), a well-established method to compress learnable weights \cite{hoefler2021sparsity}, has also been applied extensively to RSs' embedding tables \cite{pep2021}. These pruning techniques zero out unimportant embedding parameters and utilize sparse data structures to reduce the storage cost of a pruned embedding table. 
    \item \textbf{Neural architecture search} (NAS, Figure \ref{fig:related-works}.(d)) is the third mainstream type of LERS solutions. A standard NAS searches for the best model structure and hyperparameters from a predefined search space \cite{rule2021,nis2019}. In the context of LERSs, the search space is generally the embedding dimension of each feature, and the search is normally performed towards a combination of accuracy and efficiency objectives.
\end{enumerate}
We defer our discussions on the detailed methodological differences among these solutions to Section \ref{sec:related}. In addition to those three main types, there are LERS solutions that ensemble different methods to achieve embedding compression \cite{optembed2022, cerp2023}, which we term hybrid solutions.

Amid the dedication to building more advanced LERSs, it also comes to our attention that, a systematic discussion on a consistent, universal evaluation protocol for these LERSs is not yet in place in the existing literature. 
As the area of LERSs is still gradually taking its shape, these methods are often evaluated with heuristically designed protocols and a variety of benchmark datasets, which are not synchronized among publications. 
For example, as an LERS for collaborative filtering (CF) tasks, DHE \cite{dhe2021} is deployed with Generalized Matrix Factorization (GMF) and Multi-layer Perceptron (MLP) \cite{neumf2017} as recommendation backbones, and evaluated on ML-20M \cite{movielens} and Amazon-book \cite{amazonbook} datasets with Area under the ROC Curve (AUC) as the metric. In contrast, another CF solution CERP \cite{cerp2023} is evaluated on Gowalla \cite{gowalla} and Yelp2020 \cite{Yelp} benchmarks with MLP and LightGCN \cite{lightgcn2020} as backbones, where and Normalized Discounted Cumulative Gain and Recall as metrics. 
In other words, although different methods are developed for the same task, they are commonly evaluated using different datasets and metrics to benchmark performance.
Even when the same datasets and metrics are employed, the evaluation protocol can also differ in various details. 
An example is PEP \cite{pep2021}, a pruning method and OptEmbed \cite{optembed2022}, a hybrid of NAS and pruning. However, the full recommendation models compressed in \cite{pep2021} and \cite{optembed2022} differ in their original embedding dimensions (24 in \cite{pep2021} and 64 in \cite{optembed2022}), and the compressed models also vary in parameter sizes. 
These seemingly minor details can result in a significant impact on the results, as both the pre- and post-compression recommender models can reach hugely different performance when configured to different parameter sizes \cite{barsctr2021, zhang2023experimental}.


As a consequence, such inconsistency inevitably impairs the confidence when pinpointing the performance gain of the LERS proposed, and less intuitive to understand where each LERS's advantages (e.g., fast training/inference) are. As a result, this can lead to the adoption of a less effective or inappropriate solution in practical applications. 
Thus, a natural research question (RQ) arises: \textbf{(RQ1) Can we fairly benchmark a representative collection of LERSs from all categories under the same, practical evaluation setting?}

Moreover, although both collaborative filtering and content-based recommendation are arguably the most representative recommendation tasks, in LERSs, the majority of methods are designed for content-based recommendation tasks \cite{pep2021, optembed2022, qr2019} given the huge number of categorical features used to describe users and items. 
On the contrary, LERSs dedicated to collaborative filtering (CF) \cite{dhe2021,yunke_continous2023} are not intensively investigated until recently, which potentially attributes to the explosive growth of user base and item catalog in major e-commerce sites in the past few years \cite{wang2018billion,gurukar2022multibisage}. 
It is worth noting that, despite the differences in downstream outputs and recommendation models, LERSs designed for both collaborative filtering and content-based recommendation in fact bear the same goal of reducing the parameter usage of the embedding table -- one for representing content features and the other for representing user/item IDs. 
However, the embedding compression paradigms in LERSs are commonly developed in a task-specific fashion, so do the evaluations in existing papers.
Given the shared goal of embedding compression between tasks, there is another important question related to cross-task generalizability to answer: 
\textbf{(RQ2) Do methods that demonstrate strong performance in one task exhibit similarly strong performance in a different recommendation task?}

At the same time, considering that LERSs are centered around model scalability, many related metrics other than the parameter size and recommendation accuracy, especially the inference speed and runtime memory consumption, remain largely unexplored in the existing research. 
The inference speed is crucial to user experience and energy efficiency, while the runtime memory consumption is closely connected to scalability as it determines whether or not an LERS is executable on specific memory-constrained devices (e.g., TV boxes), and lower runtime memory also supports a larger batch size to speed up training.
Unfortunately, in the pursuit of lower parameter sizes, most LERSs introduce an overhead on these two metrics. For example, the compositional embedding table in TTRec \cite{ttrec2021} needs to be computed via a series of tensor multiplications, compromising the inference speed. 
Pruning-based methods, such as PEP \cite{pep2021}, introduce substantial memory overhead due to additional masks over the full embedding table.
The absence of those scalability metrics in evaluation renders it unclear if a particular LERS is a feasible solution for each given deployment configuration.
Consequently, real-world adoptions for LERSs will likely be deterred without comprehensively benchmarking their resource requirements.
Thus, we wonder: \textbf{(RQ3) How is the real-world usability of these LERSs in terms of efficiency and memory consumption during training and inference?}

Motivated to answer these questions, in this paper, we take a formal approach to benchmark various recently proposed embedding compression techniques for LERSs. 
Specifically, to address \textbf{RQ1}, we select a diversity of methods designed for collaborative filtering or content-based recommendation tasks, and performed thorough benchmarking on both tasks. For each task, we further use two real-world datasets, where all methods are universally tested with three different compression goals (i.e., the target parameter sizes after compression). For each model in each dataset, we scientifically fine-tune the hyperparameters to ensure a fair comparison.
We also point out a practically important issue -- unlike performance-oriented RS research, the field of LERSs still lacks effective baselines, that is, an easy-to-use model that provide competitive performance across different settings.
Hence, we additionally put forward \textbf{magnitude-based pruning}, a simple, effective, and performant baseline in serveral settings. Moreover, we empirically justify and analyze the suitable use cases for magnitude pruning  through our experiments.
To address \textbf{RQ2}, we extend our benchmarking across both collaborative filtering or content-based recommendation tasks for all LERSs selected, regardless of their original downstream tasks. To measure the cross-task generalizability, we first define performance retain rate, a metric to quantify how well each method preserves the full model's performance after compression. Then, we systematically analyze the obtained results to compare performance across tasks, highlighting the similarities and differences between each method's performance when being applied to the two representative recommendation tasks.
To address \textbf{RQ3}, we deploy all tested LERSs into two typical environments for LERSs: a GPU workstation and an edge device. We benchmark the time consumption and memory usage of both training and inference steps, covering both recommendation tasks. By doing so, we shed light on the overhead introduced by each method in real-world deployment. 

In summary, our main contributions are:
\begin{itemize}
    \item We extensively evaluate various LERSs' performance in two main recommendation tasks: content-based recommendation and collaborative filtering. Concretely, we cross-test different methods to verify their generalizability under different sparsity rates and tasks.
    \item We show that magnitude-based pruning, a simple baseline for embedding compression, can also achieve competitive results compared to recently proposed methods.
    \item We perform an efficiency benchmark and outline the key differences between the on-device and GPU-based settings, thus providing insights into the real-world performance of those LERSs in varying deployment environments.
    \item We release all the source codes at \href{https://github.com/chenxing1999/recsys-benchmark}{https://github.com/chenxing1999/recsys-benchmark}, which include the implementation of various embedding compression methods in PyTorch, such that the community can reuse and apply them to subsequent research problems.
\end{itemize}

\section{Related Work}\label{sec:related}
In this section, we provide relevant background for our research by reviewing the classic recommendation tasks, representative LERSs, and the benchmarking efforts in the recommendation literature. 

\subsection{Lightweight Embeddings in Recommendation}

\subsubsection{Compositional Embedding-based LERSs} 
This type of LERSs involves representing the original $n$ embedding vectors with substantially fewer parameters, where a common approach is to employ a smaller set of $m \ll n$ meta-embedding vectors. To compose a single embedding for each discrete feature, a unique subset of $t$ meta-embeddings is selected and combined. Mathematically, this is achieved by:
\begin{equation*}
\mathcal{H}(i) = \{ i_0, i_1, ..., i_t \} = \text{hash}(i),    
\end{equation*}
where $i \in \mathbb{N}$ is the original index of the feature, $\text{hash}(\cdot)$ maps $i$ into $t$ distinct indices $\{ i_1, i_2, ..., i_t \} \in \mathbb{N}_{<m}$. 
To simplify notation, let's assume that there is only one set of meta-embedding vectors, or one meta-embedding table  $\mathbf{E}^{meta}\in \mathbb{R}^{m\times d}$, then the compositional embedding $\mathbf{e}_i$ of the $i$-th feature is: 
\begin{equation*}
\mathbf{e}_i = \text{combine}(\mathbf{e}_{i_1}^{meta}, \mathbf{e}_{i_2}^{meta}, ..., \mathbf{e}_{i_t}^{meta}),
\end{equation*}
where each $\mathbf{e}_{i'}^{meta}$ corresponds to the $i'$-th row of $\mathbf{E}^{meta}$, and $\text{combine}(\cdot)$ is any operation that merges multiple vectors into one, e.g., multiplication, sum, or concatenation.
As an extension to this basic form, Shi et al. \cite{qr2019} use the quotient-remainder trick (QR) to hash the original embedding index $i$ into two new indices, which are used to extract embedding vectors from two meta-embedding tables. 
These vectors can be combined through mathematical operations, such as multiplying, adding, or concatenating, to create an embedding vector representing the original.
Following this, MEmCom \cite{memcom2020} utilizes two meta embedding tables ($\mathbf{E_1} \in \mathbb{R}^{m \times d}, \mathbf{E_2} \in \mathbb{R}^{n \times 1}$) and a pair of indices ($i_1 = i\text{ mod }m, i_2 = i$), then multiplying two meta embedding vectors to get the final embedding associated with $i$. 
The above methods are efficient as they only apply simple aggregation functions; however, this typically limits the performance due to collided meta-embeddings, especially at higher compression rates.
Another approach is TT-Rec \cite{ttrec2021}, which employs tensor-train decomposition (TTD) and a customized weight initialization to compress the embedding table. 
Because TTD can transform the exponential storage requirement into a linear function, it can achieve a higher compression rate than previous methods.
ODRec \cite{odrec2022} proposes to further compress the model with semi-tensor product-based tensor-train decomposition (STTD), and compensates for the higher compression rate with knowledge distillation. 
DHE \cite{dhe2021} proposes to use a deterministic hash function to create a pseudo-embedding table, which the authors fed to an MLP model to produce the embedding. DHE model achieves good results and can be further compressed with other techniques; however, requiring much more time to train and infer than other methods.
Based on the user-item interaction graph, LEGCF \cite{liang2024lightweight} proposes an adaptive assignment scheme through a learnable matrix, representing the weight for each meta-embedding.

\subsubsection{Pruning} This is one of the most classic approaches to compressing deep learning models in general and recommendation models in particular. These methods involve setting a portion of neuron values to zero, effectively removing them from the model. 
\begin{equation*}
\mathbf{e_i} = \hat{\mathbf{E}}_i = (\mathbf{E} \odot \mathbf{M})_i, \text{with } \mathbf{M} \in \{ 0, 1 \},
\end{equation*}
where $\mathbf{E}$ denotes the full learnable embedding table, $\mathbf{M}$ is the embedding mask, $\odot$ is element-wise multiplication. 
The main challenge of pruning-based methods is how to effectively find $\mathbf{M}$.
The most traditional approach in this category is magnitude pruning 
\cite{magnitude_pruning1989, magnitude_pruning2015}, where after the initial training phase, the weights are sorted by the magnitude of their values and the lowest-ranked ones are zeroed out. 
Leveraging this, 
DeepLight\cite{deeplight2021} gradually prunes the embedding table in the training phase based on the magnitude values until reaching the target memory budget. 
Later, PEP \cite{pep2021} applies ``soft-thresholding'' to iteratively prune the parameters in the first training step. Subsequently, the model is retrained based on the found mask with the same initialized parameters (Lottery ticket hypothesis \cite{frankle2018lottery}). 
The ``soft-thresholding'' technique allows PEP to provide a flexible embedding size for each feature.
However, PEP has high training overhead and is hard to tune for a specific compression rate due to two training steps and various hyperparameters affecting performance.
In contrast, SSEDS \cite{SSEDS} first trains the original model. Then, they determine the embedding mask with the proposed saliency score computed through the gradient and retrain the model with the newfound mask.
While SSEDS only introduces minor overhead in one forward and backward pass to calculate the embedding mask, it assumes similar embedding sizes for features in the same field, leading to constrained performance. 
Dynamic Sparse Learning (DSL) \cite{wang2024dynamic} dynamically adjusts the sparsity distribution of model weights by pruning and growth strategies to eliminate redundant parameters and activate important ones. Specifically, During training, they initially prune a significant portion of the model weights, fine-tune the model, and then prune and regrow again with a smaller amount after a few iterations.
Shaver \cite{tran2024device} employs Shapley values to efficiently and effectively prune the embedding table to any required budget for content-based recommendation models.

\subsubsection{NAS-based} Methods from this category search for the most optimal model structure in a predefined search space, typically by reinforcement learning \cite{qu2024scalable,yunke_continous2023} or an evolutionary algorithm \cite{rule2021}. They are commonly formulated as a two-level optimization problem w.r.t. both training and validation data \cite{autodim2021,zheng2024personalized}: 
\begin{equation*}
\hat{S} = \arg\!\min_S L_{val} \left( \hat{\Theta}, S \right) \text{, s.t. } \hat{{\Theta}} = \arg\!\min_{\Theta} L_{train}\left({\Theta}, S \right),
\end{equation*}
where $S$ denotes structure parameters, ${\Theta}$ denotes the model parameters.
One of the first works in this category is NIS (Neural Input Search) \cite{nis2019}, which splits the original embedding table into multiple smaller embedding blocks to create the search space. Then, NIS applies a policy network to determine the best set of embedding blocks given a memory budget.
AutoEmb \cite{autoemb2021} uses controllers that take features' popularity to suggest the embedding size of various users and items for the recommendation network. They employ differentiable architecture search (DARTS \cite{darts2018}) to solve the bi-level optimization problem, where the first and second stages are to optimize the recommendation network's weight on the training set and the controllers' weight on the validation set respectively.
While also applying DARTS, AutoDim \cite{autodim2021} uses a set of weights, which directly represent the probability of dimension sizes, and leverages the Gumbell-Softmax technique \cite{jang2017categorical} to optimize these parameters.
However, DARTS-based methods suffer from high training costs \cite{cai2018proxylessnas}.
RULE \cite{rule2021} suggests training a supernet containing various embedding blocks and an evolutionary search to search for the best embedding block set given a memory budget. To reduce the computation cost for evolutionary search, they train a performance estimator to predict an estimated performance of a given embedding block set.
CIESS \cite{yunke_continous2023} applies reinforcement learning with a random walk-based exploration strategy to efficiently identify the optimal embedding size for each user and item. 
BET \cite{bet2024} leverages a non-parametric sampler to eliminate the implicit necessity of fine-tuning a coefficient trade-off between performance and storage. This approach, however, requires multiple fine-tuning iterations of the model. To address this overhead, BET introduces a parametric performance estimator.

\subsubsection{Hybrid} \instex{Methods within this category} combine approaches from various categories. OptEmbed \cite{optembed2022} learns the pruning mask for embedding rows based on magnitude while training the supernet with uniform sampled masks for dimension sizes. Then, they apply an evolutionary algorithm to find the most optimal configuration and retrain the model with the found configuration. CERP \cite{cerp2023} integrates soft-thresholding pruning into the compositional embedding with two balanced-size embedding tables. Thus, CERP could achieve a higher compression rate than the original compositional embedding but suffers from the complexity introduced by the pruning step. \instex{ It is worth mentioning that, recent years have also seen methods approaching lightweight RSs via hardware-software co-design \cite{hsia2023mp}, but we focus our comparative study on algorithmic (software) solutions given their stronger applicability and flexibility.}


\subsubsection{Summary} The compositional methods are more straightforward to fine-tune as they define the number of parameters at the start of the training, and the training efficiency generally is better compared to other approaches. However, they suffer from limited performance because every feature has the same memory allocation, and they also introduce more inference time overhead. 
On the other hand, pruning trade-off training efficiency for a better performance \cite{zhang2023experimental}. The training pipelines are more complicated and involve multiple steps. Moreover, pruning demanded specific hardware to process sparse matrices efficiently.
NAS-based generally demands the most training resources, while having the best inference efficiency and performance.
Last but not least, hybrid methods combine the advantages of other categories to create better recommendations.

\subsection{Benchmarking Efforts in Recommender Systems}

With the increasing awareness of reproducibility in recommendation research, there has been a series of efforts in benchmarking different recommendation models. Rendle et al. \cite{rendle2019difficulty} show that well-tuned baselines could outperform newly proposed methods, which initiated a heated debate in the RSs studies. Aligning with the previous research, Maurizio et al. \cite{maurizio2019are} also indicate that simple baselines could defeat the more sophisticated deep learning models. Responding to \cite{maurizio2019are}, DaisyRec \cite{daisyrec, daisyrecv2} performs an extensive study on how various hyperparameters affect recommendation models' performance. Shehzad et al. \cite{Shehzad2023Everyone} conduct experiments to show that the worst well-fine-tuned model will outperform the best non-fine-tuned model.
With the rising concerns of reproducibility, BarsCTR \cite{barsctr2021} focuses on the reproducibility of CTR models by unifying the data pre-processing logic and providing an open-source implementation for various methods. After that, Zhu et al. \cite{bars2022} extended previous works by working on both collaborative filtering and content-based recommendation. However, these studies exclusively examine different backbones used in recommendation models. Li et al. \cite{li_survey2024} provide an overview of various LERSs methods. 
Yin et al. \cite{yin2024device} extensively explore the on-device settings for RSs, including inference, training, and security concerns.
Zhang et al. \cite{zhang2023experimental} share similarities with our work, studying various embedding compression methods, albeit with the limited scope of recommendation task (CTR prediction only) and minimal hyperparameter fine-tuning. On the other hand, our work provides a more extensive hyperparameter tuning by adapting the methodology from \cite{daisyrecv2} and studies LERSs' transferability onto collaborative filtering tasks.



\section{Base Recommenders}
The LERSs we have selected for benchmarking are compatible with the majority of base recommenders, as long as they are latent factor models with an embedding layer. To ensure fairness, all LERSs are plugged into the same base recommender in every test. We evaluate two representative recommendation tasks, specifically the content-based recommendation and collaborative filtering. 
This section describes the backbones selected for the two tasks, namely  NeuMF \cite{neumf2017} and LightGCN \cite{lightgcn2020,simgcl2022} for collaborative filtering, and DeepFM \cite{deepfm2017} and DCN\_Mix \cite{dcnv2} for content-based recommendation.

\subsection{Collaborative Filtering}

The most classic approach for RSs is collaborative filtering, which solely takes user-item past interactions to produce user-item affinity. 
In our experiments, users' interests are represented by a binary value to indicate whether an interaction occurred (implicit feedback).
For testing LERSs on the collaborative filtering task, we adopt two commonly used backbones: a latent-factor-based model NeuMF \cite{neumf2017} and a graph-based model LightGCN \cite{lightgcn2020}. 
We will further elaborate on these methods in the section below.

\subsubsection{NeuMF}
The latent-factor-based collaborative filtering models' target is to find the shared latent representation for users and items from the interaction matrix \cite{marcuzzo2022recommendation}.
One of the most well-established in this category is matrix factorization, which associates each user and item with an embedding vector. 
Let $\mathbf{e}_u$ and $\mathbf{e}_v$ denote the embedding vector for user $u$ and item $v$, respectively. Matrix factorization computes the relevant score $r_{uv}$ as the dot product between $\mathbf{e}_u$ and $\mathbf{e}_v$. 
Later, NeuMF \cite{neumf2017} proposes generalized matrix factorization (GMF), which modifies the original dot product formula by adding learnable parameters $\mathbf{h}$, and combines it with the DNN branch to further enhance matrix factorization. The outputs of these two branches are defined as:

\begin{equation*}
    \begin{aligned}
    \hat{y}^{\text{GMF}}_{uv} &= \mathbf{h}^\top \left( \mathbf{e}^{\text{GMF}}_u \odot \mathbf{e}^{\text{GMF}}_v \right), \\
    \hat{y}^{\text{DNN}}_{uv} &= \text{DNN} \left( \mathbf{e}^{\text{DNN}}_u, \mathbf{e}^{\text{DNN}}_v\right),
    \end{aligned}
\end{equation*}
where $\mathbf{e}^\text{GMF}_{\cdot}$ and $\mathbf{e}^\text{DNN}_{\cdot}$ is the embedding corresponding for GMF and DNN branch, respectively. $\odot$ denotes element-wise product of vectors. It is noteworthy that these two embedding vectors are different. The final score is computed by summing two branches' results:
\begin{equation*}
    r_{uv} = \sigma \left( \hat{y}^{\text{GMF}}_{uv} +  \hat{y}^{\text{DNN}}_{uv} \right),
\end{equation*}

where $\sigma \left( \cdot \right)$ is the sigmoid function. Finally, based on the original paper, we optimize the model with log loss function and an $L_2$ penalty, as defined in the following:
\begin{equation*}
    L_{NeuMF} = - \sum_{(u, v^+, \mathcal{V}^-) \in \mathcal{B}} \!\! \left( \text{ln } \left( r_{uv^+} \right) + \sum_{v^- \in \mathcal{V}^-}\text{ln } \left( 1 - r_{uv^-} \right) \right) + \lambda\lVert \Theta \rVert^2,
\end{equation*}
where each training batch $\mathcal{B}$ consists of training samples $(u, v^+, \mathcal{V}^-)$ that are constructed by pairing a user $u$ with one of her interacted item $v^+$ and a set $\mathcal{V}^-$ of uninteracted items. Following the common practice \cite{daisyrecv2,cerp2023}, we also apply an $L_2$ regularization on all learnable parameters $\Theta$ with weight $\lambda$ to prevent overfitting.

\subsubsection{LightGCN}
For graph-based collaborative filtering, the user-item interactions are formulated as a bipartite graph, where an edge between user $u$ and item $v$ nodes exists if $u$ has an observed interaction with item $v$. Taking a user node $u$ as an example, we denote $\mathcal{N}(u)$ as the set of $u$'s one-hop neighbors. At each layer $l$, the node embedding is updated by aggregating embeddings from all the neighbors within $\mathcal{N}(u)$:
\begin{equation*}
    \mathbf{e}_u^{(l)} = \sum_{z \in \mathcal{N}(u)} (|\mathcal{N}(u)|\cdot |\mathcal{N}(z)|)^{-\frac{1}{2}} \mathbf{e}_u^{(l-1)},
\end{equation*}
where $\mathbf{e}_u^{(l)}$ is the embedding of $u$ in the $l$-th layer, ${(|\mathcal{N}(u)|\cdot|\mathcal{N}(z)|})^{-\frac{1}{2}}$ is the normalization term. Notably, when $l=0$, $\mathbf{e}_u^{(0)} $ corresponds to the $u$-th row of the embedding table $\mathbf{E}$, which is randomly initialized and learned via back-propagation. After $L$ layers' propagation, the final representation of user $u$ is obtained by averaging the embeddings from all layers:
\begin{equation*}
\mathbf{e}_{u} = \frac{1}{L+1} \sum_{l=0}^L \mathbf{e}_u^{(l)},
\end{equation*}
where the embedding $\mathbf{e}_v$ for an arbitrary item $v$ is obtained analogously. 
To facilitate ranking, the relevance score $r_{uv}$ for the user-item pair is the dot product between their final embeddings, i.e., $r_{uv} = \langle\mathbf{e}_{u}, \mathbf{e}_{v} \rangle$. 
The loss function employed in the original paper \cite{lightgcn2020} combines the Bayesian Personalized Ranking (BPR) loss and $L_2$ penalty, as defined in the following:
\begin{equation*}
    L_{BPR} = - \sum_{(u, v^+, v^-) \in \mathcal{B}} \!\! \text{ln } \sigma \left( r_{uv^+} - r_{uv^-} \right) + \lambda \lVert \mathbf{E} \rVert^2,
\end{equation*}
where each training batch $\mathcal{B}$ consists of training samples $(u, v^+, v^-)$ that are constructed by pairing a user $u$ with one of her interacted item $v^+$ and an unvisited one $v^-$. As the trainable parameters in LightGCN only contain the embedding table, the $L_2$ regularization is only enforced on $\mathbf{E}$ with weight $\lambda$. However, on larger datasets, Yu et al. \cite{simgcl2022} point out that LightGCN is prone to prolonged convergence time. To address this issue, it is recommended to incorporate InfoNCE \cite{infonce2018} into the loss function of LightGCN:
\begin{equation*}
    L_{LightGCN} = L_{BPR} - \gamma \sum_{z \in \mathcal{B}} \text{ln} \frac{\text{exp}(1 / \tau)}{\sum_{z' \in \mathcal{B}} \text{exp} (\langle\mathbf{e}_z,\mathbf{e}_{z'}\rangle/\tau)},
\end{equation*}
where $\gamma$ is the weight for the InfoNCE loss, $\tau$ is the temperature, and $z$ and $z'$ are users/items in the sampled batch $\mathcal{B}$. Note that the above InfoNCE requires no data augmentation for the sake of efficiency \cite{simgcl2022}, hence the softmax directly uses $1 / \tau$ as the numerator. As proven in \cite{simgcl2022}, the addition of the InfoNCE loss effectively improves the quality of embeddings learned in every epoch and substantially reduces the convergence time\footnote{There is an approximately $50\times$ speed-up in our observation, hence we resort to this configuration throughout the benchmarking.} as a result.

\subsection{Content-based Recommendation}
In this section, we first introduce the content-based recommendation task, then two backbones we used in our benchmark.
In content-based recommendation, we have $n$ features encoded in vector $\mathbf{x} \in \mathbb{R}^n$, where $x_i \in \mathbf{x}$ represents the value of feature $i$. Taking CTR prediction as an example, $\mathbf{x}$ is commonly the concatenation of the features of a user and an advertisement, where the recommender is expected to estimate the probability of a click behavior. For the ground truth $y$ to be predicted, $y=1$ if a click is observed, and $y=0$ otherwise. Considering the features in $\mathbf{x}$ are commonly sparse \cite{fm2010}, it is essential to learn effective feature interactions. To this end, the factorization machine (FM) \cite{fm2010} has been a well-established approach for content-based recommendation: 
\begin{equation*}
    \hat{y}_{FM} = w_0 +\sum_{i=1}^{n}w_{i}x_i +\sum_{i=1}^{n}\sum_{j=i+1}^{n}\langle {\mathbf{e}}_{i}, {\mathbf{e}_{j}}\rangle \cdot x_ix_j,
\end{equation*}
where $\langle \cdot, \cdot \rangle$ is the dot product, and $w_0$ is the bias term to be learned. For each feature $i$, $w_i$ is the learnable scalar weight, and $\mathbf{e}_i \in \mathbb{R}^d$ is its corresponding embedding drawn from the embedding table $\mathbf{E}$. For model optimization, we follow the common practice \cite{alpt2023,optembed2022,deeplight2021} and adopt the following log loss: 
\begin{equation*}
    L_{CTR}=- y\ln \hat{y}_{FM} - (1-y)\ln(1-\hat{y}_{FM}) + \lambda\lVert \Theta \rVert^2,
\end{equation*}
which quantifies the prediction error between $\hat{y}$ and the ground truth $y$. In addition, the second term denotes the $L_2$ regularization over all trainable parameters $\Theta$ to prevent overfitting, and $\lambda$ is a tunable coefficient that controls its weight in the loss function.

Despite the versatility of FM, it only accounts for second-order feature interactions, hence being  insufficient for the more complex applications. 
Therefore, various methods were proposed to model higher-order feature interactions.

\subsubsection{DeepFM} \cite{deepfm2017} proposes to uplift the expressiveness of FM by incorporating a deep neural network (DNN) branch into the original FM model to model the higher-order interactions between features:
\begin{equation*}
   \hat{y}_{DNN} = \text{DNN} \Big{(}\text{pool}(\{x_i\mathbf{e}_{i}\}_{i=1}^n)\Big{)},
\end{equation*}
with the pooling operator $\text{pool}(\cdot)$. In DeepFM, the pooling operation is done by firstly performing sum pooling over the weighted embeddings $x_i\mathbf{e}_{i}$ in each feature field (e.g., user region and movie genre), and then concatenating the results from all fields \cite{deepfm2017}. The final prediction is calculated as an ensemble:
\begin{equation*}
    \hat{y} = \sigma \left( \hat{y}_{FM} + \hat{y}_{DNN} \right),
\end{equation*}
where $\sigma(\cdot)$ is the sigmoid function. Notably, the feature embeddings used in both FM and DNN branches are drawn from the same embedding table $\mathbf{E}$.

\subsubsection{DCN--Mix} 
DCNv2 \cite{dcnv2} explicitly models $\left( l + 1 \right)$-level interaction with $l$ cross layers:
\begin{equation}
    \mathbf{e}^{\left( l+1 \right)} = \mathbf{e}^{\left( 0 \right)} \odot \left( \mathbf{W}^{\left( l \right)} \mathbf{e}^{\left( l \right)} + \mathbf{b}^{\left( l \right)} \right) + \mathbf{e}^{\left( l \right)},
    \label{eq:dcn-cross}
\end{equation}
where $\mathbf{e}^{\left( 0 \right)} \in \mathbb{R}^f $ is the result of pooling operator $\text{pool}(\{x_i\mathbf{e}_{i}\}_{i=1}^n)$. $\mathbf{e}^{\left( l \right)}, \mathbf{e}^{\left( l+1 \right)}$ represents the input and output of the (l+1)-th cross layer respectively. $\mathbf{W}^{\left( l \right)} \in \mathbb{R}^{f \times f}$ and $\mathbf{b}^{\left( l \right)} \in \mathbb{R}^f$ is the learnable weight matrix and bias vector for the corresponding $l$-th layer.  To reduce the training cost, we employ the DCN--Mix backbone. 
First, Wang et al. \cite{dcnv2} suggest to utilize low-rank approximation to reduce the compute cost of Eq. \ref{eq:dcn-cross}:

\begin{equation*}
    \mathbf{e}^{\left( l + 1 \right)} = \mathbf{e}^{\left( 0 \right)} \odot \left( \mathbf{U}^{\left( l \right)} \left( \mathbf{V}^{\left( l \right)\top} \mathbf{e}^{\left( l \right)} \right) + \mathbf{b}^{\left( l \right)} \right) + \mathbf{e}^{\left( l \right)},
\end{equation*}
where $\mathbf{U}^{\left( l \right)}, \mathbf{V}^{\left( l \right)} \in \mathbb{R}^{f \times r}$ and rank $r \ll f$. Based on this intuition, DCN--Mix applied the idea of Mixture-of-Experts (MoE), which consists of two modules: experts (typically small models, denoted as $P_i: \mathbb{R}^f \rightarrow \mathbb{R}^f$) and gating (a function $G_i: \mathbb{R}^f \rightarrow \mathbb{R}$). Each expert specializes in handling a specific data distribution, while the gating network assigns a weight to each expert, determining how much each expert should contribute to the final output. The specific computation is provided below:
 
\begin{equation*}
\begin{aligned}
\mathbf{e}^{\left( l + 1 \right)} &= \sum_{i=1}^{K} G_i \left( \mathbf{e}^{\left( l \right)} \right) P_i  \left( \mathbf{e}^{\left( l \right)} \right) + \mathbf{e}^{\left( l \right)}, \\
P_i(\mathbf{e}^{\left( l \right)}) &= \mathbf{e}^{\left( 0 \right) } \odot \left( 
    \mathbf{U}_i^{\left( l \right)} \cdot 
        \text{tanh} \left( \mathbf{C}_i^{\left( l \right)} \cdot 
        \text{tanh} \left( \mathbf{V}^{\left( l \right)\top}_{i}  \mathbf{e}^{\left( l \right)} \right) \right) 
        + \mathbf{b}^{\left( l \right)} 
\right), \\
G_i (\mathbf{e}^{\left( l \right)}) &= \mathbf{W}_i \mathbf{e}^{\left( l \right)}
\end{aligned}
\end{equation*}
where $K$ is the number of experts, $ \mathbf{U}_i^{\left( l \right)},  \mathbf{V}^{\left( l \right)}_{i} \in \mathbb{R}^{f \times r}$ and $\mathbf{C}_i^{\left( l \right)} \in \mathbb{R}^{r \times r}$ is the learnable parameters of $i$-th expert's $l$-th layer, $\mathbf{W}_i \in \mathbb{R}^{f \times 1}$ is the learnable parameters of $i$-th gate, tanh is the non-linear activation function.
Finally, as $L$ cross layers can only model upto $\left( L + 1 \right)$ interaction orders, the authors propose incorporating a deep neural network to further enhance modeling capacity. Specifically, we adopt the proposed stacked structure, which works empirically better for the Criteo dataset \cite{dcnv2}.
The specific procedure is defined below:

\begin{equation*}
    \hat{y} = \sigma \left( \text{DNN} \Big{(}\mathbf{e}^{\left( L \right)} \Big{)} \right),
\end{equation*}
where $\mathbf{e}^{\left( L \right)}$ is the output of the last cross layer.

\subsection{Tree-structured Parzen Estimator (TPE) for Hyperparameter Tuning}
\label{sec:TPE}


\instex{To ensure the fairness of comparisons, hyperparameter tuning is an indispensable step. Grid search is the most basic hyperparameter optimization algorithm, which tests each set of hyperparameters based on a predefined grid. However, grid search has an exponential cost with the number of hyperparameters and cannot be applied on a continuous range, thus limiting its performance. However, the vast combinatorial space of tunable hyperparameters in most embedding compression methods introduces a major challenge in terms of efficiency. Therefore, in this work, we employ Tree-structured Parzen Estimator (TPE) \cite{tpe2011} as our hyperparameter optimization algorithm, which is a common choice in the recommender system literature \cite{dcnv2,daisyrec,daisyrecv2}. In what follows, we introduce how TPE is leveraged for hyperparameter tuning.}

\instex{In a nutshell, TPE attempts to estimate $P(\mathcal{X}|y)$, where $\mathcal{X}$ is a set of hyperparameters, and $y$ is the associated performance measured by AUC (for CTR prediction tasks) or NDCG (for collaborative filtering tasks). 
Specifically, TPE maintains two distributions $P(\mathcal{X}|y > y^*)$ and $P(\mathcal{X}|y \leq y^*)$ with $y^*$ being a predefined performance threshold. 
These two distributions are modeled by two Gaussian Mixture models.
Initially, we randomly run each method with $\omega$ ($\omega=10$ in our case) sets of hyperparameters, and $y^*$ is selected based on the most performant setting. Then, based on results from the initial $\omega$ trials, we select the next hyperparameter setting $\mathcal{X}$ such that it maximizes the expected ratio $\frac{P(\mathcal{X}| y > y^*)}{P(\mathcal{X} | y \leq y*)}$. After finishing a trial, both $P(\mathcal{X}|y > y^*)$ and $P(\mathcal{X}|y \leq y^*)$ are updated according to the sampled $\mathcal{X}$ and its evaluation result. When performing hyperparameter tuning for each method, we repeat the described process until a maximum number of $30$ trials (including the initial $\omega$ trials) has been reached, and use the setting $\mathcal{X}$ that yields the best validation performance for testing.}


\section{LERS Baselines Compared}
In this section, we first introduce the LERSs methods used in our benchmark. 
Then, we propose a straightforward baseline for comparative analysis.

\subsection{Chosen Baselines}
\begin{table}[]
    \centering

         

\caption{Summary of chosen LERSs with a high-level description. \textbf{P}, \textbf{C}, and \textbf{N} denote Pruning-, Composition-, and NAS-based methods, respectively. Section 4.1 provides further discussions on each method.}
\begin{tabular}{@{}l|ll|ccc|p{0.6\textwidth}}
\hline
Method & Venue & Year & P & C & N & Description \\ \hline
QR \cite{qr2019} & KDD & 2020 & & $\bullet$ & &
\begin{tabular}[c]{@{}p{0.6\textwidth}@{}} \instex{
To compute an embedding vector, the original index is hashed into two codes based on the quotient-remainder trick. Then, two meta-embeddings are retrieved based on the hash codes, and are combined using element-wise multiplication to represent the target feature.}
\end{tabular} \\ \hline
TTRec \cite{ttrec2021} & MLSys & 2021 & & $\bullet$ & & \begin{tabular}[c]{@{}p{0.6\textwidth}@{}}\instex{The original embedding table is split into multiple smaller tensors, called TT-cores. Then, to retrieve an embedding vector, it indexes and multiply entries across these TT-cores sequentially.}\end{tabular} \\ \hline
DHE \cite{dhe2021} & KDD & 2021 & & $\bullet$ & & \begin{tabular}[c]{@{}p{0.6\textwidth}@{}}
\instex{
    It employs a heuristic-based, deterministic hash function to convert each item index into a dense embedding vector. Then, we pass the resulting embedding vector through a trainable MLP to obtain the final embedding.
}
\end{tabular} \\ \hline
PEP \cite{pep2021} & ICLR & 2021 & $\bullet$ & & & \begin{tabular}[c]{@{}p{0.6\textwidth}@{}}
    \instex{Through reparameterization, a soft threshold can be learned as a cut-off point for pruning embedding table entries. Then, by only keeping the retained embedding entries, the recommendation model is retrained from scratch.
}\end{tabular} \\ \hline
OptEmb \cite{optembed2022} & CIKM & 2022 & $\bullet$ & & $\bullet$ & \begin{tabular}[c]{@{}p{0.6\textwidth}@{}}
    \instex{First, a super-network consisting of multiple embedding configuration is trained. Then, an evolutionary search algorithm is used to select the best sub-network structure from the super-network, which is then retrained.}
\end{tabular} \\ \hline
CERP \cite{cerp2023} & ICDM & 2023 & $\bullet$ & $\bullet$ & & \begin{tabular}[c]{@{}p{0.6\textwidth}@{}}
    \instex{It integrates trainable threshold into two balanced meta-embedding tables. To compensate for sparse embedding tables, an additional regularization loss is proposed to encourage complementary pruning of the two meta-embedding tables.}
\end{tabular} \\ \hline
\end{tabular}
    \vspace{-.5cm}
    \label{tab:emb_comp}
\end{table}

We select these techniques based on two primary aspects: \textbf{influence} (should be from a renowned venue or highly cited) and \textbf{diversity} (should cover all main LERS types).
Table \ref{tab:emb_comp} provides a summary of our chosen methods and their categories.

\subsubsection{QR} 
    Shi et al. \cite{qr2019} divided the original embedding table $\mathbf{E} \in \mathbb{R}^{n \times d}$ into two new tables, $\mathbf{E}_1 \in \mathbb{R}^{p \times d}, \mathbf{E}_2 \in \mathbb{R}^{q \times d}$, where $p$ is a hyperparameter and $q = \lceil n / p \rceil $.  
    To retrieve the embedding for the original index $i$, we use two new indices $i_1=i \text{ mod } p$ and $i_2=i \text{ div } p$ to extract embedding vectors from $\mathbf{E}_1$ and $\mathbf{E}_2$.
    Finally, these vectors can be combined through various mathematical operations. In our experiments, we employ element-wise product between two vectors as it is the most competitive method in the original paper.
    We chose QR as it is one of the most simple compositional embedding methods while being widely accepted as a baseline and inspired various other methods \cite{cerp2023,coleman2023unified}.

\subsubsection{TTRec}
    Yin et al. \cite{ttrec2021} proposed a novel algorithm to compress RSs embedding based on tensor-train decomposition (TT). TTRec factorizes the number of items $n \leq \prod_{i=1}^t n_i$ and the hidden size $d \leq \prod_{i=1}^t d_i$ into integers, and decomposes the embedding table by $\mathbf{E} \approx \mathbf{G}_1 \mathbf{G}_2 \dots \mathbf{G}_t$, where TT-core $\mathbf{G}_i \in \mathbb{R}^{r_{i-1} \times n_i \times d_i \times r_i}$. To further improve the efficiency, they introduced a cache mechanism and leveraged the cuBLAS (CUDA Basic Linear Algebra Subroutine) library. 
    We selected TTRec as the representative for a more complex compositional embedding method. TTRec also inspired various other works \cite{odrec2022,nimblett2022}. \instex{Notably, compared with QR which only reduces the feature dimension of embedding table ($n$), TTRec reduces both the feature dimension ($n$) and the hidden dimension ($d$) via tensor decomposition.}
    
\subsubsection{DHE}
    The core idea of DHE \cite{dhe2021} is utilizing a hash function to transform the original index $i$ into a new dense vector $\mathbf{v} \in \mathbb{R}^{k}$, with $k$ being a large number (for example, 1024).
    This hash function is deterministic, thus requiring almost no storage cost.
    They then feed this vector $\mathbf{v}$ into an MLP model, whose number of parameters is much smaller than the original embedding table, to produce the actual embedding.
    We chose this method due to its initial design for CF and its proven effectiveness in compressing item and user embeddings.

\subsubsection{PEP}
    PEP \cite{pep2021} draws inspiration from Soft Thresholding Reparameterization (STR) \cite{str2020}, which gradually prunes the model by learning threshold $\mathbf{s}$ through backpropagation.
    This threshold determines which parameters to prune dynamically.
    STR pushes small weights towards zero while preserving significant ones, allowing the model to retain important features and maintain performance. 
    Finally, PEP retrained the model from scratch with the same parameter initialization with the found pruning mask.
    PEP also influences various other pruning methods \cite{SSEDS,optembed2022,cerp2023,wang2024dynamic}, generally regarding finding the winning lottery ticket.

\subsubsection{OptEmbed}
    OptEmbed \cite{optembed2022} considers pruning models as two tasks separately: Which feature to keep -- represented by a binary mask $\mathbf{m}_e$, and what is the dimension size should be allocated to each field -- represented by an integer array $\mathbf{m}_d$. 
    The model is trained with three separate steps. The first step is to train a supernet model and find $\mathbf{m}_e$, while possible values for $\mathbf{m}_d$ are sampled from uniform distribution and incorporated into the training procedure. 
    In the second step, OptEmbed performs an evolutionary search to find the best settings for $\mathbf{m}_d$. Finally, they retrained the model from scratch with the same parameter initialization with the found pruning mask.
    We chose OptEmbed due to its better training efficiency compared to other NAS-based approaches.
    
\subsubsection{CERP}
    CERP \cite{cerp2023} integrates STR into two balanced embedding tables $\mathbf{E}_1$ and $\mathbf{E}_2$.
    To compensate for two sparse meta-embedding tables, the authors incorporate a regularization loss and opt for vector summation as the combination operation, thus creating a dense embedding vector from two sparse meta-embedding vectors.
    CERP is a prime example of a hybrid approach between the two most common LERSs: compositional embedding and pruning.

\subsection{Magnitude-based Pruning as An Intuitive Baseline}
In this section, we describe magnitude-based pruning (MagPrune), a pruning method we have proposed for lightweight embeddings. MagPrune is a simple, intuitive, and easy-to-deploy baseline in a wide range of recommendation tasks. 
We also provide the rationale for its working mechanism in embedding pruning.

To be specific, for either the content-based recommendation or collaborative filtering task, we first train the base recommender with $L_2$ regularization. Given the formulation of $L_2$, it restrains the magnitude for model parameters, and further encourages a lower magnitude for less vital parameters. As such, for the learned embedding table $\mathbf{E}$, we sort all entries' absolute values and set the lower value to zeros until the memory budget is satisfied. This process is defined in Algorithm \ref{algo:mag-prune}.

\begin{algorithm}[t!]
\caption{Magnitude-based Pruning (MagPrune)}\label{euclid}
\label{algo:mag-prune}
\begin{algorithmic}[1]
\Procedure{MagPrune}{$t,n_{min},\mathbf{E}$}
\State $num\_element \gets \mathbf{E}.shape[0] * \mathbf{E}.shape[1]$
\State $num\_prune \gets num\_element * t$
\State $\mathbf{\hat{E}} \gets \lVert \mathbf{E} \rVert$ \Comment{Calculate absolute value by element-wise}
\State $topk\_indices = get\_topk(\mathbf{\hat{E}}, k=n_{min}, dim=1)$ \Comment{Get $n_{min}$ largest elements for each row}
\State $\mathbf{\hat{E}}\left[ topk\_indices \right] \gets \infty$

\State $indices \gets argsort(\mathbf{\hat{E}}.flatten())$
\State $\mathbf{E}[indices[:num\_prune]] \gets 0$
\State \textbf{return} $\mathbf{E}$
\EndProcedure
\end{algorithmic}
\end{algorithm}

In the above steps, $t\in (0\%,100\%]$ is a specified target compression ratio, and $n_{min}$ is the minimum number of parameters allocated to each embedding row. For the CTR task, $n_{min}$ will be set to 0, while for the CF task, $n_{min}$ will be searched by choosing the value that optimizes the validation NDCG. This hyperparameter is introduced with the intuition that each user and item should have at least one parameter representing their information. In what follows, we justify the reason why MagPrune is able to be competent as a baseline for embedding pruning. 

\begin{figure}[t!]
    \centering
    \begin{subfigure}[b]{0.23\textwidth}
        \centering
        \includegraphics[width=\textwidth,height=\textheight,keepaspectratio]{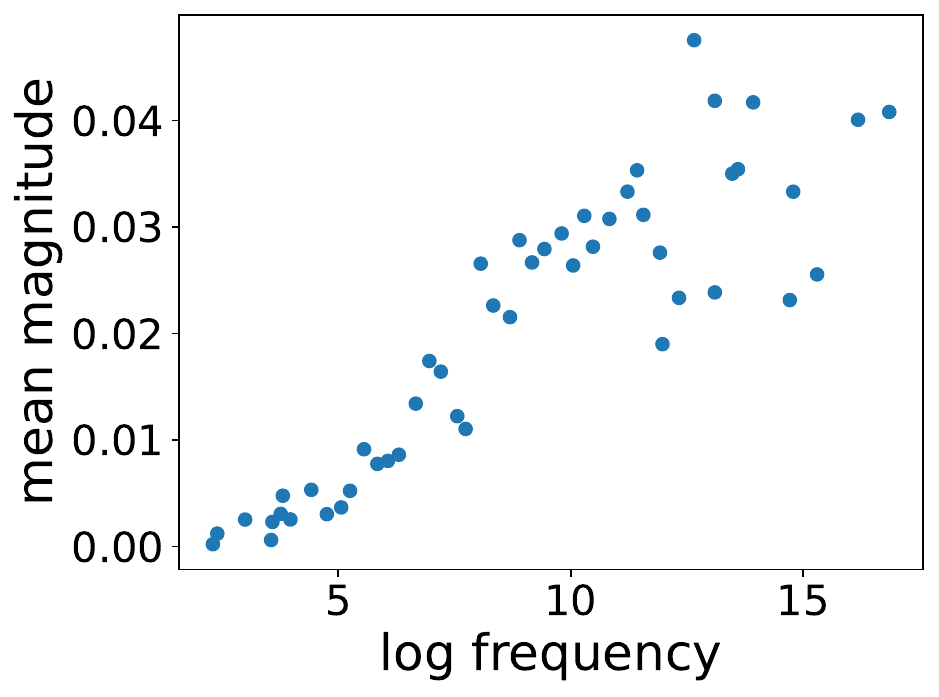}
        \vspace{-0.6cm}
        \caption{Field 1 features in Criteo}
        \label{fig:log-freq-mag-criteo1}
    \end{subfigure}
    \hfill
    \begin{subfigure}[b]{0.23\textwidth}
        \centering
        \includegraphics[width=\textwidth,height=\textheight,keepaspectratio]{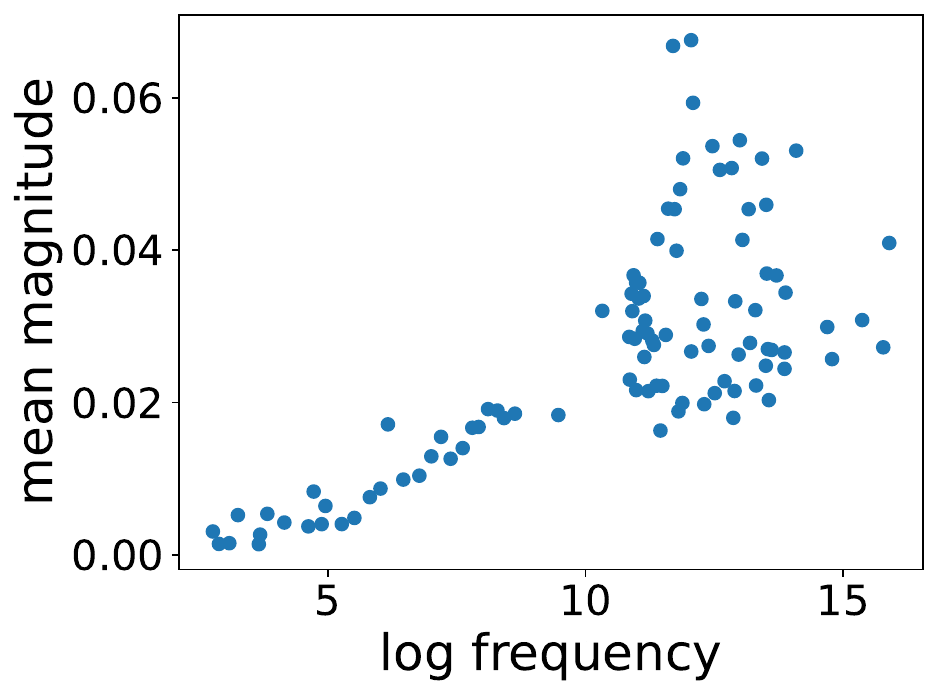}
        \vspace{-0.6cm}
        \caption{Field 2 features in Criteo}
        \label{fig:log-freq-mag-criteo2}
    \end{subfigure}
    \begin{subfigure}[b]{0.23\textwidth}
        \centering
        \includegraphics[width=\textwidth,height=\textheight,keepaspectratio]{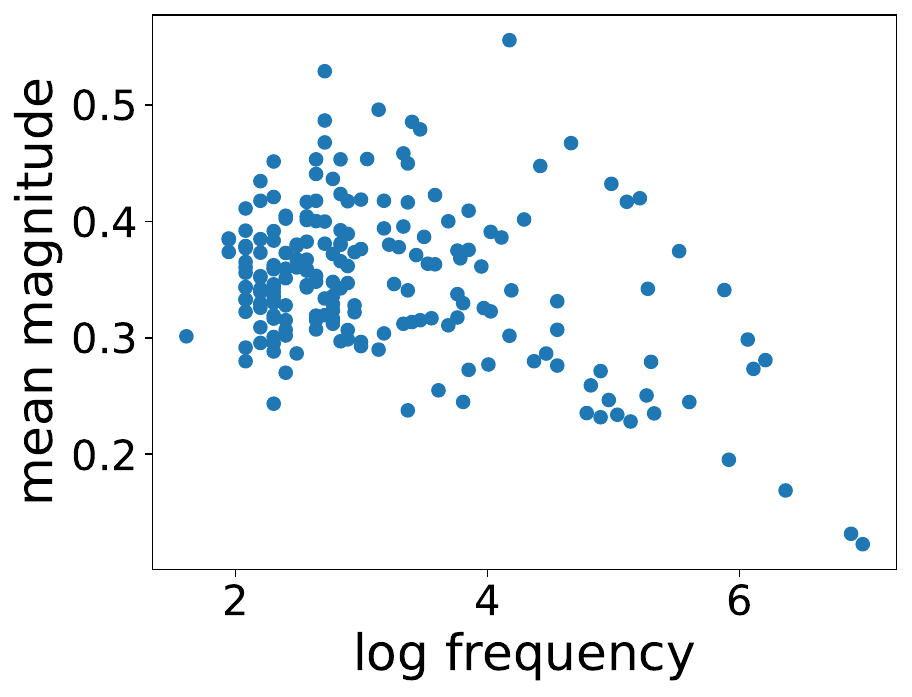}
        \vspace{-0.6cm}
        \caption{Items in Gowalla}
        \label{fig:log-freq-mag-yelp}
    \end{subfigure}
    \hfill
    \begin{subfigure}[b]{0.23\textwidth}
        \centering
        \includegraphics[width=\textwidth,height=\textheight,keepaspectratio]{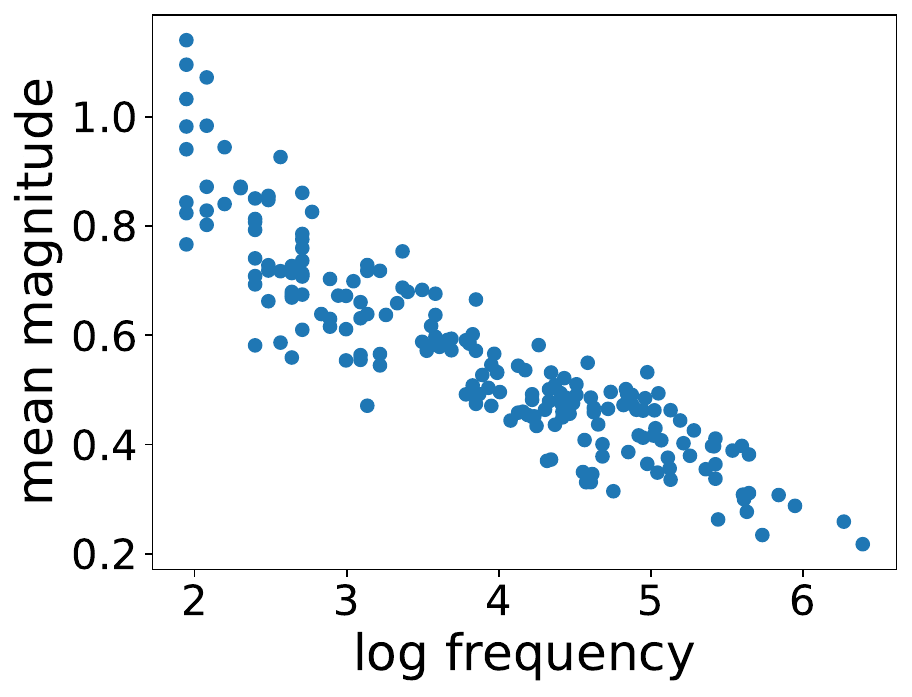}
        \vspace{-0.6cm}
        \caption{Users in Gowalla}
        \label{fig:log-freq-mag-gowalla}
    \end{subfigure}
    \caption{\instex{Log frequency of features and their respective mean magnitude in the learned embeddings. Dataset details are provided in Section \ref{sec:exp_setting}. We visualize all features in the first and second feature fields in Criteo. For Gowalla, we limit the user/item number to 200 for better visibility.}}
    \label{fig:log-freq-mag}
    \vspace{-0.3cm}
\end{figure}

In Figure \ref{fig:log-freq-mag}, we plot the relationship between the frequency of discrete features and the mean absolute magnitude of their embeddings learned with $L_2$ regularizer. 
In the content-based recommendation dataset Criteo (Figure \ref{fig:log-freq-mag-criteo1} and \ref{fig:log-freq-mag-criteo2}), embeddings with higher frequency correspond to the common features, whose embedding magnitude appears to be higher than that of the low-frequency features.
\instex{This phenomenon depicts how the $L_2$ and recommender systems' embedding table work together. In short, low-frequency features have lower contributions to the main recommendation loss, leading to zero or near-zero gradients for their embeddings. As a result, their embeddings are predominantly influenced by gradients from the $L_2$ term, which keeps most of their values within a small range.}
This aligns with a general observation \cite{autoemb2021, pep2021} that the high-frequency features appear in more training samples, and their embeddings are usually more informative and dominant when generating predictions. Thus, MagPrune can effectively take out the less informative dimensions of low-frequency feature embeddings. 

Interestingly, in the collaborative filtering dataset Gowalla (Figure \ref{fig:log-freq-mag-yelp} and \ref{fig:log-freq-mag-gowalla}), due to the nature of the graph-based model LightGCN, a different trend is observed as the high-frequency features (i.e., popular users and items) tend to have a lower average magnitude in their learned embeddings. In this occasion, pruning the embedding table by maintaining high-magnitude embedding dimensions means that, more emphasis is laid on users/items that are in the mid or tail range of the long-tail distribution. Furthermore, as the user/item representations in Gowalla are learned by LightGCN, popular users/items are in fact high-degree nodes in the interaction graph with abundant neighbors, hence being able to mostly depend on their neighbors' embeddings to form their own representations. 

\instex{Generally, for learned model weights, an absolute value close to zero commonly indicates less contribution to the model outputs. Following this intuition, we 
can remove most parameters with low magnitudes.} 
In Section \ref{sec:exp}, we will further demonstrate the competitive recommendation accuracy achieved by MagPrune. At the same time, MagPrune provides strong efficiency advantages due to its capability of fitting multiple target compression ratios after a one-off training cycle.

\section{Experiment Settings}\label{sec:exp_setting}




To answer the three questions raised in Section \ref{sec:intro}, we have designed a series of experiments to benchmark the performance and efficiency of the selected LERSs. In this section, we present our experimental settings in detail.  

\begin{table}[b]
    \centering
    \caption{Statistics of the preprocessed datasets for content-based recommendation (CTR prediction).}
    \vspace{-0.3cm}
    \begin{tabular}{l|rrr}
    \hline
         Dataset  & \#Instances & \#Features & \#Fields  \\
         \cline{1-4}
         Criteo & $45,840,617$ & $1,086,810$ & $39$ \\
         Avazu & $40,428,967$ & $4,428,511$ & $22$\\
    \hline
    \end{tabular}
    \label{tab:ctr_dataset_stat}
\end{table}

\subsection{Datasets}

For the content-based recommendation (CTR prediction) task, we utilize two widely used benchmark datasets: \textbf{Criteo} \footnote{https://www.kaggle.com/c/criteo-display-ad-challenge} and \textbf{Avazu} \footnote{https://www.kaggle.com/c/avazu-ctr-prediction/data} \cite{pep2021,optembed2022}. The data pre-processing logic is based on \cite{barsctr2021}. Table \ref{tab:ctr_dataset_stat} shows the core statistics of processed datasets. For Criteo, we apply the optimal solution from the Criteo contest, where we discretize each value $x$ in numeric feature fields to $\ceil{\text{log}_2(x)} $ if $x > 2$. For Avazu, we remove the `ID' feature, which is unique for every sample and not useful for the CTR task. Then, for both datasets, we replace infrequent features (appearing less than 10 times in Criteo and 2 times in Avazu) as out-of-vocabulary (OOV) tokens. For both two datasets, we randomly split them into 8:1:1 as the training, validation, and test sets respectively.

For the collaborative filtering (top-$k$ recommendation) task, we utilize two datasets, \textbf{Gowalla} and \textbf{Yelp2018}, that are widely adopted in the literature \cite{lightgcn2020,ngcf2019}. The main statistics are presented in Table \ref{tab:cf_dataset_stat}. To enhance reproducibility, we directly employ the data split provided by LightGCN \cite{lightgcn2020}. As \cite{lightgcn2020} did not supply a validation set, we generate our own training and validation sets from the original training data. Specifically, we divide each user's interactions in the training set with a 9-to-1 ratio, where the former part is used for training and the latter is for validation.

\begin{table}[t!]
    \centering
    \caption{Statistics of the preprocessed datasets for collaborative filtering (top-$k$ recommendation).}
    \vspace{-0.3cm}
    \begin{tabular}{l|rrrr}
    \hline
         Dataset  & \#User & \#Item & \#Interactions & Sparsity  \\
         \cline{1-5}
         Yelp2018 & $31,668$ & $38,048$ & $1,561,406$ & 99.87\% \\
         Gowalla & $29,858$ & $40,981$ & $1,027,370$ & 99.92\%\\
    \hline
    \end{tabular}
    \label{tab:cf_dataset_stat}
\end{table}

\subsection{Evaluation Protocols}
Each method is tested under three sparsity rates $t$ 50\% (low sparsity), 80\% (moderate sparsity), and 95\% (high sparsity). 
To calculate the sparsity rate, we only consider the embedding table parameters instead of all trainable parameters. 
The sparsity rate $t$ is defined as follows: 
\begin{equation*}
    t = 1 - \frac{M ( \hat{\mathbf{E}} ) }{M( \mathbf{E} )},
\end{equation*}
where the memory function $M \left( \cdot \right)$ outputs the number parameters to construct embedding table. 
We implemented different functions to calculate parameters to suit each compression method.

\subsubsection{Evaluating Recommendation Accuracy}
We adopt different evaluation metrics for CTR prediction and top-$k$ recommendation tasks, which are introduced below. 
\begin{itemize}
    \item For CTR prediction, we evaluate all models with LogLoss and AUC (area under the ROC curve). 
AUC measures the probability that a random positive example has a higher probability than a random negative example. In RSs, an improvement of 0.001 in AUC is generally considered significant \cite{marcuzzo2022recommendation,barsctr2021}.
LogLoss measures the difference between the predicted likelihood of a click and the actual outcome.
It is worth noting that AUC only considers ranking, while LogLoss takes into the exact output value. The higher AUC indicates better RSs, while lower LogLoss indicates better performance.
Each compression method is applied once to the single general embedding table instead of separately for each field. Based on the settings from DaisyRec \cite{daisyrecv2}, we train each base configuration for 30 trials, with 15 epochs for each trial with AUC as the target metric.
We choose the best configuration based on the validation set from those $30\times 15$ checkpoints.

    \item For top-$k$ recommendation, we adopt the commonly used ranking metric NDCG$@k$ and Recall$@k$ with $k=20$. Normalized Discounted Cumulative Gain (NDCG) evaluates the quality of a top-$k$ recommendation by comparing the ranked relevance of recommended items to the ideal ranking. Recall quantifies the proportion of relevant items successfully retrieved out of the total relevant ones available in the test set.  In the evaluation step of top-$k$ recommendation, we first calculated scores for each user and pair, then removed all existing pairs in the training set. Finally, for each user, we take the top $k$ scored items.
    Each compression method, except for the pruning-based ones, is applied individually to user and item embedding tables separately. The exception is due to the difficulty in attaining the target sparsity for both embedding tables simultaneously in pruning-based scenarios. Similarly to CTR prediction, we train each base configuration for 30 trials, with 40 epochs for each. 
Similarly, the best configuration is chosen based on the validation performance from all checkpoints. 
\end{itemize}

\subsubsection{OptEmbed \cite{optembed2022} Modification}  In all CF settings, we remove feature mask $\mathbf{m}_e$, which chooses rows to prune, thus keeping every user and item embedding vectors.
Because the authors didn't design the method to reach a flexible memory, besides providing the original method's performance, we modify the distribution of sampling $\mathbf{m}_d$  in evolutionary step to increase sparsity rate when required. 

Specially, Lyu et al. \cite{optembed2022} use a uniform distribution to sample the embedding dimension mask $\mathbf{m}_d$, which results in the expected sparsity rate of roughly 50\%. To increase sparsity rates, we modified the distribution of OptEmbed based on \cite{petrov2022effective}. The modified distribution is defined as follows:
\begin{equation*}
    p_i = \frac{\alpha^{h -i}}{\sum_{i=1}^h \alpha^{h-i}},
\end{equation*} 
where $p_i$ is probability of sampled embedding size $i$, $h$ is maximum hidden size, $\alpha$ is a hyperparameter to control the distribution. If $\alpha=1$, the above distribution is equal to the original uniform distribution used by OptEmbed \cite{optembed2022}.
We have following equations:
$$
\sum_{i=1}^h \alpha^{h-i} = \frac{\alpha^h - 1}{\alpha - 1},
$$
$$
\sum^h_{i=1} \left( {\alpha^{h - i} \times i} \right) = \frac{h - (h+1) \alpha + \alpha^{h+1}}{(\alpha - 1)^2}.
$$

The expected hidden size is:
\begin{equation*}
    \begin{split}
            \mathbb{E}(d) &= \sum^h_{i=1} p_i \times i \\
            &= \frac{\sum^h_{i=1} \left( {\alpha^{h - i} \times i} \right)}{\sum_{i=1}^h \alpha^{h-i}} \\ 
            &= \frac{h - (h+1) \alpha + \alpha^{h+1}}{(\alpha - 1) (\alpha^h - 1)}, \\
            \mathbb{E}(d) &= \frac{\alpha}{\alpha - 1} - \frac{h}{\alpha^h - 1}, \text{with } \alpha \neq 1.
    \end{split}
\end{equation*}

With the above formula, we could approximate $\alpha$ with gradient descent and mean squared error. Note that for the original uniform distribution, $\mathbb{E}(d) = (h + 1) / 2$, which approximated the $50\%$ sparsity rate ($h/2$). Then, we sample from the proposed distribution and only take results with a sparsity higher than the target sparsity. 

In 80\% of Criteo -- DeepFM and Avazu -- DCN pairs, we keep the original uniform distribution but only take candidates with sparsity higher than the target because the original model already has a close sparsity rate with the target.
We only use our method to increase sparsity and don't train the model with lower sparsity than the original models.

\subsubsection{Implementation Details}
\instex{We hereby provide an additional note on the hyperparameter tuning process: 
\begin{itemize}
    \item For compositional-based methods (TTRec, QR and DHE), we first find a hyperparameter configuration that satisfies the parameter budget constraint. Then, we keep modifying the hyperparameters within the parameter budget (e.g., adding more MLP layers with reduced width in DHE). Finally, for each parameter budget, we choose the best configuration for every dataset-backbone combination.
    \item For other methods, it is trickier to control the trade-off between accuracy and model size. Generally, for these methods (PEP, OptEmbed and CERP), we use the existing configuration reported by the authors if possible. If not, we focus on tuning the hyperparameters specific to each method, e.g., the soft threshold initialization in PEP.
\end{itemize} 
After the optimal method-specific hyperparameters are tuned, we finetune the general hyperparameters, such as learning rate and weight decay. This step is done through TPE as per Section \ref{sec:TPE}, which is a widely used hyperparameter optimization approach \cite{dcnv2, daisyrecv2}.}

\subsubsection{Evaluating Cross-task Transferability}

To evaluate each method's cross-task transferability, we calculate the performance retain ratio between the compressed model and the original model:
\begin{equation*}
    \text{performance retain ratio} = \frac{\text{compressed model performance metric}}{\text{original model performance metric}}.
\end{equation*}

In our experiments, we employ NDCG as the evaluation metric for CF and use LightGCN as the model backbone due to its superior performance. 
For the CTR task, we use AUC as the evaluation metric and DCN as the backbone, selected for its improved performance compared to DeepFM.
Then, we compute the overall performance retain ratio by averaging the results between two datasets. 

\subsubsection{Evaluating Real-world Efficiency}

For efficiency, we aim to benchmark two main metrics -- runtime and peak memory of training and inference phase on two devices, namely a GPU workstation and a Raspberry Pi which respectively mimic the deployment environments of a GPU server and a smaller edge device. 
\begin{itemize}
    \item The GPU workstation uses an i7-13700K CPU, NVIDIA RTX A5000 GPU, and 32GB RAM. 
    \item The Raspberry Pi is the 4B generation with quad-core Cortex-A72 (ARM v8) and 4GB SDRAM.
\end{itemize}

We implemented most methods using Python and PyTorch High-level API. 
For deployment, we will utilize the corresponding compiled version of Pytorch (the CUDA version for workstations and the ARM version for Raspberry Pi).
For methods that cannot be with PyTorch high-level API, namely TTRec and accessing elements in sparse matrices, we provided the Numba and CUDA implementation.
We deploy all LERSs in both environments, and record the runtime during both training and inference, as well as the peak VRAM (for GPU) and RAM (for Raspberry Pi) usage.
Based on device capacity, the batch sizes in use are 2,048 for the workstation, and 64 for the Raspberry Pi.
For edge device training benchmarks, we only ran 20\% and 5\% of an epoch for CF and CTR, respectively, then linearly scaled the runtime accordingly.
We use DeepFM as backbone for CTR prediction and LightGCN as backbone for top-$k$ recommendation.
In the CTR task, the runtime is measured as the time needed to get the predicted score $\hat{y}$. 
In top-$k$ recommendation, the runtime is measured as the necessary time to compute all user and item embeddings. This is because the runtime for recovering a full user/item embedding from the reduced embedding parameters (1.87ms) is substantially more significant than calculating the user-item similarity via dot product (0.041ms) and ranking (0.1ms).

We have benchmarked all methods under the moderate 80\% sparsity rate configuration. In this work, we adopt the conventional SparseCSR format to store sparse matrices, acknowledging that the overhead of sparse matrices depends on the level of sparsity \cite{hoefler2021sparsity}. 

\begin{table}[t!]
\caption{Hyperparameter settings. LightGCN and NeuMF correspond to the base models for collaborative filtering, while DeepFM and DCN correspond to content-based recommendation tasks. The hyperparameters are searched with the TPE \cite{tpe2011} algorithm.}
\vspace{-0.5cm}
\renewcommand{\arraystretch}{0.9}
\setlength\tabcolsep{3pt}
\center
  \begin{tabular}{c c c c}
    \toprule
    \multirow{2}{*}{Base Model} & Hyper- & \multirow{1}{*}{Fix} & \multirow{2}{*}{Search Interval}\\
    & parameter & Value &\\
    \hline
    \multirow{5}{*}{LightGCN}
      & $\tau$ & 0.2 & - \\
      & $d$ & 64 & - \\
      & learning rate & - & \text{[} $5 \times 10^{-4}$, $10^{-2}$ \text{]} \\
	 & $\lambda$ & - & \text{[} $10^{-5}$, $10^{-2}$ \text{]}  \\
	 & $L$ & - & \{1,2,3,4\} \\
	 & $\gamma$ & - & [0,1] \\
    \hline
        \multirow{5}{*}{NeuMF}
      & $MLP$ & $64, 32, 16$ & - \\
      & $d$ & 32 & - \\
      & learning rate & - & \text{[} $5 \times 10^{-4}$, $10^{-2}$ \text{]} \\
      & $\lambda$ & - & \text{[} $10^{-5}$, $10^{-2}$ \text{]}  \\
	 & dropout & - & \text{[} 0, 1 \text{]}  \\
	 & $\lVert \mathcal{V}^- \rVert$ & - & \{1,2,3,4,5\} \\
    \hline
    \multirow{5}{*}{DeepFM} 
      & MLP & $400, 400, 400$ & - \\
	 & $d$ & $16$ & - \\
	 & dropout & - & \text{[} 0, 1 \text{]} \\
	 & learning rate & - & \text{[} $10^{-5}, 10^{-2}$ \text{]} \\
	 & $\lambda$ & - & \text{[} $10^{-5}, 10^{-2}$ \text{]} \\
      \hline
    \multirow{7}{*}{DCN--Mix}
      & MLP & $512, 512$ & - \\
      & $r$ & $64$ & - \\
      & $L$ & $3$ & - \\
      & $K$ & $4$ & - \\
	 & $d$ & $16$ & - \\
	 & dropout & - & \text{[} 0, 1 \text{]} \\
	 & learning rate & - & \text{[} $10^{-5}, 10^{-2}$ \text{]} \\
	 & $\lambda$ & - & \text{[} $10^{-5}, 10^{-2}$ \text{]} \\

	 \bottomrule
\end{tabular}
\label{table:hyperparams}
\vspace{-0.3cm}
\end{table}

\begin{table*}[t!]
    \centering
\renewcommand{\arraystretch}{1.2}
\setlength\tabcolsep{0.72pt}
\caption{Results on CTR prediction with DeepFM and DCN as the base models. The sparsity $t$ indicates the number of pruned parameters (the higher $t$ is, the more parameters are reduced). \#Para indicates the parameter size of the embedding table (M: million, K: thousand), and Loss refers to the LogLoss. In each column, the best result is marked in bold and the second best one is underlined.}
\vspace{-0.2cm}
\begin{tabular}{cl|llr|llr||llr|llr}
\hline
    & & \multicolumn{6}{c||}{Criteo} & \multicolumn{6}{c}{Avazu} \\
\cline{3-14}
    & & \multicolumn{3}{c|}{DeepFM} & \multicolumn{3}{c||}{DCN} & \multicolumn{3}{c|}{DeepFM} & \multicolumn{3}{c}{DCN} \\
\cline{3-14}
     Sparsity    & Method &  AUC      & Loss & \#Para  &  AUC      & Loss & \#Para  & AUC      & Loss & \#Para &  AUC      & Loss & \#Para \\
\cline{1-14}
    0\% & Original & 0.8102 & 0.4416 & 17.39M & 0.8114 & 0.4407 & 17.39M 
                   & 0.7658 & 0.3932 & 70.86M & 0.7759 & 0.3839 & 70.86M \\
\cline{1-14}
 -\% & OptEmb & 0.8088 & 0.4429 & 4.41M & 0.8112 & 0.4408 & 1.45M 
                & 0.7585 & 0.3936 & 3.41M & 0.7671 & 0.3878 & 19.81M \\
\cline{1-14}
\multirow{6}{*}{$\sim 50\%$} 
& QR\cite{qr2019}       & 0.8081  & 0.4435 & 8.69M & 0.8095 & 0.4425 & 8.69M 
                        & \textbf{0.7697} & \textbf{0.3864} & 35.43M & 0.7743 & 0.3857 & 35.43M           \\
& TTRec\cite{ttrec2021} & 0.8075  & 0.4442 & 9.36M & 0.8112 & 0.4408 & 8.26M
                        & 0.7647  & 0.3915 & 35.43M & 0.7711 & 0.3873 & 35.43M         \\
& DHE\cite{dhe2021}     &  0.8050  & 0.4467 & 8.69M & 0.8095 & 0.4427 & 8.69M
                        & 0.7554 &  0.3959 & 31.74M & 0.7594 & 0.3921 & 31.74M         \\
& PEP\cite{pep2021}     & \textbf{0.8105}       & \textbf{0.4414} & 8.67M & \textbf{0.8113} & \textbf{0.4407} & 8.69M
                        & 0.7647 & 0.3886  & 29.67M  & \textbf{ 0.7754} & \underline{0.3852} & 33.55M      \\
& CERP\cite{cerp2023}   & 0.8099  & 0.4419  & 8.66M  & 0.8110 & 0.4410 & 8.66M       
                        & 0.7649 & \underline{0.3885} & 35.30M & 0.7723 & \underline{0.3852} & 35.34M     \\
& MagPrune              & \underline{0.8102}  & \underline{0.4417} & 8.69M & \textbf{0.8113} & \textbf{0.4407} & 8.69M
                        & \underline{0.7655}  & 0.3932 & 35.43M & \textbf{0.7754} & \textbf{0.3841} & 35.43M          \\
\cline{1-14}
\multirow{7}{*}{$\sim 80\%$} 
& QR\cite{qr2019}       & 0.8078       & 0.4438 & 3.48M & 0.8091 & 0.4427 & 3.48M
                        & \textbf{0.7695} & \textbf{0.3871} & 14.17M & 0.7698 & 0.3922 & 14.17M \\
& TTRec\cite{ttrec2021} & 0.8070       & 0.4446 & 3.15M & 0.8103 & 0.4416 & 3.48M
                        & \underline{0.7651} & 0.3902 & 14.17M & \underline{0.7710} & 0.3900 & 14.17M            \\
& DHE\cite{dhe2021}     & 0.8053       & 0.4463 & 3.17M & 0.8109 & 0.4412 & 3.17M
                        & 0.7536 & 0.3957 & 14.74M & 0.7650 & 0.3916 & 14.74M           \\
& PEP\cite{pep2021}  & \underline{0.8098} & \underline{0.4419} & 3.47M & 0.8108 & 0.4413 & 3.46M
                         & 0.7633  & 0.3896 & 14.07M & 0.7669 & 0.3906 & 12.57M           \\
& OptEmb\cite{optembed2022}     & 0.8088 & 0.4430 & 3.17M & \textbf{0.8112} & \textbf{0.4408} & 1.45M 
                        & 0.7585 & 0.3936 & 3.41M & 0.7663 & 0.3877 & 13.24M             \\
& CERP\cite{cerp2023}   & 0.8095 & 0.4423 & 3.21M & 0.8106 & 0.4416 & 3.44M
                        & 0.7638 & \underline{0.3892} & 14.14M & 0.7673 & \underline{0.3873} & 13.05M          \\
& MagPrune              & \textbf{0.8101} &\textbf{ 0.4417} & 3.48M & \underline{0.8111} & \underline{0.4410} & 3.48M
                         & 0.7604  & 0.3962 & 14.17M & \textbf{0.7713} & \textbf{0.3862} & 14.17M           \\
\cline{1-14}   
\multirow{7}{*}{$\sim 95\%$} 
& QR\cite{qr2019}       & 0.8033  & 0.4482 & 870K & 0.8064 & 0.4452 & 870K
                        & \textbf{0.7644}    & \textbf{0.3905} & 3.54M & 0.7652 & \textbf{0.3895} & 3.54M           \\
& TTRec\cite{ttrec2021} & \textbf{0.8087} & \textbf{0.4431}  & 870K & 0.8106 & 0.4414 & 870K
                        & \underline{0.7608} & 0.3924 &  3.54M & \underline{0.7678} & 0.3903 &  3.54M         \\
& DHE\cite{dhe2021}     & 0.8050 & 0.4466 & 846K & 0.8098 & 0.4420 & 846K 
                        & 0.7601     & 0.3964 & 3.35M & 0.7677 & 0.3913 &  3.35M          \\
& PEP\cite{pep2021}     & \underline{0.8084}       & \underline{0.4432}   & 806K & \textbf{0.8114} & \textbf{0.4406} & 831K  
                        & 0.7590     & 0.3922 & 3.50M & 0.7564 & 0.3938 & 3.52M             \\
& OptEmb\cite{optembed2022}      & 0.8043 & 0.4471 & 863K  & \underline{0.8109} & \underline{0.4408} & 856K
                        & 0.7585 & 0.3936 & 3.41M & \textbf{0.7683} & \textbf{0.3895} & 3.43M  \\
& CERP\cite{cerp2023}   & 0.8062  & 0.4454  & 864K & 0.8107 & 0.4414 & 850K
                        & 0.7607 & \underline{0.3920} & 3.54M & 0.7631 & 0.3933 & 3.54M          \\
& MagPrune              & 0.8080        & 0.4436  & 869K & 0.8086 & 0.4434 & 869K 
                        & 0.7501    & 0.4034 & 3.54M & 0.7602 & 0.3924 & 3.54M        \\
\hline 
\end{tabular}
    \label{tab:deepfm-exp-result}
\end{table*}

\begin{table*}[t!]
    \centering
\renewcommand{\arraystretch}{1.2}
\setlength\tabcolsep{1.3pt}
\caption{Results on top-$k$ recommendation with LightGCN and NeuMF as the base models. In each column, the best result is marked in bold and the the second best one is underlined.}
\vspace{-0.2cm}
\begin{tabular}{cl|llr|llr||llr|llr}
\hline
     & & \multicolumn{6}{c||}{Yelp2018} & \multicolumn{6}{c}{Gowalla} \\
\cline{3-14} 
     & & \multicolumn{3}{c|}{LightGCN} & \multicolumn{3}{c||}{NeuMF} & \multicolumn{3}{c|}{LightGCN} & \multicolumn{3}{c}{NeuMF} \\
    \cline{3-14} 
        Sparsity    & Method & N@20      & R@20 & \#Para      & N@20      & R@20 & \#Para  & N@20      & R@20 & \#Para  & N@20      & R@20 & \#Para \\
\cline{1-14}
    0\%       & Original & 0.0575       & 0.0727 & 4.46M  & 0.0348 & 0.0451 & 4.46M      
              & 0.1470      & 0.1787 & 4.53M & 0.1066 & 0.1304 & 4.53M            \\
\cline{1-14}
\multirow{7}{*}{$\sim 50\%$} 
& QR\cite{qr2019}       & 0.0523       & 0.0663 & 2.23M & 0.0287 & 0.0372 & 2.23M
                        & 0.1380      & 0.1674 & 2.27M & 0.0775 & 0.0944 & 2.27M          \\
& TTRec\cite{ttrec2021} & 0.0513       & 0.0652 & 2.30M & 0.0272 & 0.0348 & 2.05M
                        &  0.1372      & 0.1651 & 2.20M & 0.0747 & 0.0918 & 2.35M            \\
& DHE\cite{dhe2021}     & 0.0516      & 0.0661 & 2.17M & 0.0267 & 0.0353 & 1.88M
                        &  0.1375      & 0.1672 & 2.17M & 0.0691 & 0.0868 & 1.88M            \\
& PEP\cite{pep2021}     & \underline{0.0560} & \underline{0.0712} & 2.21M 
                        & \underline{0.0306} & \underline{0.0398} &	2.22M  
            &  \underline{0.1431} & \underline{0.1741} & 2.24M 
            & \textbf{0.0959} & \textbf{0.1193} & 2.26M    \\
& OptEmb\cite{optembed2022}     & 0.0499       & 0.0634 & 2.27M  &\textbf{0.0320} & \textbf{0.0415} & 2.30M 
                        &  0.1338      & 0.1605 & 2.31M & 0.0890 & 0.1066 & 2.33M           \\
& CERP\cite{cerp2023}   & 0.0506       & 0.0641 & 2.22M & 0.0237 & 0.0304 & 2.22M
                        & 0.1376       & 0.1659 & 2.26M & 0.0675 & 0.0814 & 2.26M           \\
& MagPrune           & \textbf{0.0566} & \textbf{0.0718}  & 2.23M  & 0.0297 & 0.0388 & 2.23M
             & \textbf{0.1447}       & \textbf{0.1761} & 2.27M 
             & \underline{0.0949} & \underline{0.1165} & 2.27M \\

\cline{1-14}
\multirow{7}{*}{$\sim 80\%$} 
& QR\cite{qr2019}       & 0.0443       & 0.0559 & 893K & 0.0259 & 0.0334 & 893K  
                        & 0.1229       & 0.1485 & 907K & 0.0654 & 0.0752 & 907K    \\
& TTRec\cite{ttrec2021} & 0.0442       & 0.0562 & 896K & 0.0225 & 0.0291 & 816K
                        & 0.1245       & 0.1499 & 890K & 0.0648 & 0.0784 & 930K     \\
& DHE\cite{dhe2021}     & 0.0495       & 0.0627 & 775K & 0.0258 & 0.0338 & 887K
                        & \textbf{0.1300}   & \textbf{0.1573} & 775K & 0.0654 & 0.0789 & 887K \\
& PEP\cite{pep2021}  & \textbf{0.0539} & \textbf{0.0682}  & 866K & \textbf{0.0295} & \textbf{0.0382} & 892K
        & \underline{0.1284}       & \underline{0.1561} & 823K & \textbf{0.0913} & \textbf{0.1097} & 907K \\
& OptEmb\cite{optembed2022}     & 0.0403       & 0.0513 & 886K & \underline{0.0270} & \underline{0.0345} & 877K
                    & 0.1168       & 0.1403 & 899K & \underline{0.0679} & \underline{0.0797} & 905K           \\
& CERP\cite{cerp2023}   & 0.0423       & 0.0535 & 864K & 0.0231 & 0.0296 & 891K
                        & 0.1171       & 0.1421 & 858K & 0.0613 & 0.0736 & 906K        \\
& MagPrune              & \underline{0.0497}       & \underline{0.0633} & 892K & 0.0187 & 0.0188 & 892K
                        & 0.1257       & 0.1528 & 907K & 0.0558 & 0.0683 & 907K           \\
\cline{1-14}
\multirow{7}{*}{$\sim 95\%$} 
& QR\cite{qr2019}       & 0.0347 & 0.0442 & 226K & 0.0194 & 0.0255 & 226K
                        & 0.1011  & 0.1232 & 229K & 0.0500 & 0.0605 & 229K \\
& TTRec\cite{ttrec2021} & 0.0364      & 0.0464 & 202K & 0.0156 & 0.0200 & 234K
                        & 0.1076      & 0.1305 & 229K & 0.0458 & 0.0558 & 239K           \\
& DHE\cite{dhe2021}     & \textbf{0.0400}      & \textbf{0.0515} & 227K & 0.0184 & 0.0239 & 233K
                        &  \underline{0.1112}     & 0.1322 & 227K & 0.0508 & 0.0636 & 233K            \\
& PEP\cite{pep2021}  & \underline{0.0396}  & \underline{0.0504} & 223K & \textbf{0.0285} & \textbf{0.0363} & 223K
                     &    \textbf{0.1156} & \textbf{0.1410} & 226K & \textbf{0.0816} & \textbf{0.0957} & 227K            \\
& OptEmb\cite{optembed2022}     & 0.0287 & 0.0365 & 222K & 0.0187 & 0.0234 & 198K
                     &  0.0772 & 0.0913 & 226K & 0.0475 & 0.0522 & 200K             \\
& CERP\cite{cerp2023}   & 0.0357 & 0.0453 & 223K & \underline{0.0204} & \underline{0.0256} & 223K
                        & 0.1109 & \underline{0.1331} & 227K & \underline{0.0536} & \underline{0.0654} & 227K	            \\
& MagPrune              & 0.0197 & 0.0244 & 223K & 0.0091 & 0.0111 & 223K
                        & 0.0365 & 0.0474 & 227K & 0.0261 & 0.0321 & 227K             \\
\hline                     
\end{tabular}
    \label{tab:lightgcn-exp-result}
\end{table*}

\section{Analysis on Experimental Results}\label{sec:exp}
In this section, we discuss the results of the experiments conducted using the aforementioned settings. 

\subsection{Overall CTR Performance (RQ1)}


Table \ref{tab:deepfm-exp-result} shows the CTR prediction task experiment results. 
 In line with \cite{zhang2023experimental, coleman2023unified}, we observe that for the Criteo dataset, the pruning methods usually outperform others, while in the Avazu dataset, the compositional encoding-based methods outperform others. Coleman et al. \cite{coleman2023unified} hypothesize that the Criteo dataset has a heavier tail feature distribution than the Avazu dataset, consequently, having more colliding tokens and worsening the errors.
However, we could see that the log loss of pruning methods in the Avazu dataset is lower than compositional encoding-based methods.
Interestingly, PEP and MagPrune perform similarly at a low sparsity rate across three out of four pairs of dataset and backbone. This similarity indicates the strong performance of MagPrune, especially in low sparsity rates.
Additionally, the difference is minimal between the original model and the worst performance in 95\% compression rate (less than 1\% for Criteo and 3\% for Avazu for both backbones).

For the Criteo dataset, 
PEP demonstrated the most competitive performance relative to other methods across two backbones and three sparsity rates, only failing for DCN at the medium sparsity rate. 
With only 5\% of the parameters, PEP could achieve the performance of the original DCN model, further highlighting PEP's competitiveness on the Criteo dataset. 
As mentioned above, the pruning methods generally outperform the compositional-based methods. 
For example, in the 50\% sparsity rate, both PEP and MagPrune achieved the highest performance for both backbones. 
OptEmbed relative performance for DCN is higher than DeepFM, as shown at 80\% and 95\% sparsity rates. It appears that OptEmbed is more suitable with the DCN backbone compared to DeepFM.  

For the Avazu dataset, we would argue that TTRec is generally a good choice for the high sparsity rate, while MagPrune and QR are for low sparsity. 
First, TTRec consistently demonstrated impressive performance, being the second-best for both backbones. 
Similarly, in the low sparsity rate, both QR and MagPrune achieve the top three in AUC for both backbones. 
Regarding the fluctuation in the relative performance between sparsity rates, one potential explanation is that at lower sparsity rates, removing parameters has less catastrophic impact than at higher sparsity rates.
Thus, simple compression methods under proper tuning already achieved competitive performance at lower sparsity rates.
While higher sparsity rates, where the parameters are much less, require complex methods to maintain the performance.

By taking a deep look at each method, we could see that TTRec and DHE could have performance gains when allocated fewer parameters.
Furthermore, CERP performs competitively despite being initially developed for CF tasks.
An interesting note is that we observe no performance gain for OptEmbed on the validation set through the retraining step of the CTR prediction in most experiment settings (except for DCN -- Avazu).
In this task, magnitude pruning shows competitive results even at a high compression rate.

\subsection{Overall Collaborative Filtering Performance (RQ1)}

Table \ref{tab:lightgcn-exp-result} shows the CF task experiment results.
PEP delivers competitive results in every setting. In general, similar to the CTR task, simple methods (QR, MagPrune) perform better at low sparsity rates, while more complex methods (TTRec, CERP) perform better at higher sparsity rates. 
As mentioned above, we hypothesize that complex model compression methods preserve essential features more effectively, which becomes vital as sparsity increases, thus maintaining better performance in higher sparsity rates. 
On the other hand, simple methods perform well at lower sparsity rates, where the impact of removed parameters is less pronounced, making their straightforward approach sufficient.
Compared to the CF task, where the relative performance depends more on the dataset side, the relative performance in the CF task depends more on backbones, as NeuMF differs from LightGCN more greatly compared to DCN and DeepFM.
Additionally, LightGCN consistently outperforms NeuMF across all settings, highlighting the robust performance of graph-based models in CF in general and LERS in particular.


For NeuMF, PEP outperforms all the other methods significantly in medium and high sparsity rates while still having competitive performance in low sparsity rates, consistently in the top 2. 
In the medium sparsity rate, the second best is OptEmbed; however, it cannot maintain its performance in the high sparsity rate. 
The performance of MagPrune already significantly dropped as it cannot leverage the graph structure.
In the high sparsity rate, the second best is CERP -- the method specifically catered to high sparsity scenarios.
Interestingly, for NeuMF, the gap between the best method and the others is much more notable.

For the LightGCN backbone, simple MagPrune achieves the best results at a moderate compression rate, followed by PEP, which shows that the fine-grain pruning-based method achieves high performance at various compression rates. 
Additionally, DHE also demonstrated competitive performance with LightGCN.
In contrast, OptEmbed assigns zero to the latter dimensions of most embedding, leading to zero embedding in most latter dimensions with LightGCN's propagation method. 
Moreover, during our hyperparameter tuning process for OptEmbed 95\% for both datasets, the best hyperparameter's InfoNCE loss coefficient is 0, as they don't converge with InfoNCE loss, further limiting their performance.
This has led to poor performance of OptEmbed for LightGCN backbone.

\subsection{Cross-task Transferability (RQ2)}

\begin{table}[t]
\caption{Performance retain ratio between the original and the compressed models. In each column, the best result is marked in bold and the second best one is underlined.}

\begin{tabular}{ll|rr|r||rr|r}
\hline
Sparsity & Method & Yelp2018        & Gowalla       & Overall         & Criteo         & Avazu & Overall \\
\hline
\multirow{7}{*}{$\sim 50\%$} 
& QR & 0.9096 & 0.9388 & 0.9242 & 0.9977 & 0.9979 & 0.9978 \\
& TTRec & 0.8922 & 0.9333 & 0.9128 & 0.9998 & 0.9938 & 0.9968 \\
& DHE & 0.8974 & 0.9354 & 0.9164 & 0.9977 & 0.9787 & 0.9882 \\
& PEP & 0.9739 & 0.9735 & \underline{0.9737} & 0.9999 & 0.9994 & \textbf{0.9996} \\
& OptEmb & 0.8678 & 0.9102 & 0.8890 & 0.0000 & 0.0000 & 0.0000 \\
& CERP & 0.8800 & 0.9361 & 0.9080 & 0.9995 & 0.9954 & 0.9974 \\
& MagPrune & 0.9843 & 0.9844 & \textbf{0.9844} & 0.9999 & 0.9994 & \textbf{0.9996} \\
\hline

\multirow{7}{*}{$\sim 80\%$} 
& QR & 0.7704 & 0.8361 & 0.8032 & 0.9972 & 0.9921 & 0.9947 \\
& TTRec & 0.7687 & 0.8469 & 0.8078 & 0.9986 & 0.9937 & \underline{0.9962} \\
& DHE & 0.8609 & 0.8844 & \underline{0.8726} & 0.9994 & 0.9860 & 0.9927 \\
& PEP & 0.9374 & 0.8735 & \textbf{0.9054} & 0.9993 & 0.9884 & 0.9938 \\
& OptEmb & 0.7009 & 0.7946 & 0.7477 & 0.9998 & 0.9876 & 0.9937 \\
& CERP & 0.7357 & 0.7966 & 0.7661 & 0.9990 & 0.9889 & 0.9940 \\
& MagPrune & 0.8643 & 0.8551 & 0.8597 & 0.9996 & 0.9941 & \textbf{0.9969} \\
\hline

\multirow{7}{*}{$\sim 95\%$} 
& QR & 0.6035 & 0.6878 & 0.6456 & 0.9938 & 0.9862 & 0.9900 \\
& TTRec & 0.6330 & 0.7320 & 0.6825 & 0.9990 & 0.9896 & \underline{0.9943} \\
& DHE & 0.6957 & 0.7565 & \underline{0.7261} & 0.9980 & 0.9894 & 0.9937 \\
& PEP & 0.6887 & 0.7864 & \textbf{0.7375} & 1.0000 & 0.9749 & 0.9874 \\
& OptEmb & 0.4991 & 0.5252 & 0.5122 & 0.9994 & 0.9902 & \textbf{0.9948} \\
& CERP & 0.6209 & 0.7544 & 0.6876 & 0.9991 & 0.9835 & 0.9913 \\
& MagPrune & 0.3426 & 0.2483 & 0.2955 & 0.9965 & 0.9798 & 0.9882 \\
\hline

\end{tabular}
\label{tab:cross-task}
\end{table}

Table \ref{tab:cross-task} shows the performance ratio between the compressed model and the original model. 
First, we could observe that both PEP and MagPrune consistently perform well in two recommendation tasks when the sparsity rate is low.
At higher sparsity rates of the CF task, PEP maintains its strong performance, followed by DHE, a method designed for collaborative filtering.
DHE performs less competitively in the CTR task, especially in lower sparsity rate.
This might be because a neural network with the same parameter size has a much stronger representation capability than the traditional embedding table, thus making DHE more prone to overfit in a higher parameter budget.
In contrast, OptEmbed demonstrates strong results on CTR but performs poorly on CF.
As mentioned above, OptEmbed is not suitable for the LightGCN model.

For the CTR task, MagPrune still performs well at the medium sparsity rate but fails slightly short compared to other methods at the high sparsity rate.
Surprisingly, at the high sparsity rate, PEP performs best for LightGCN but is the worst for DCN. 
TTRec appears to be a more consistent method for this task, especially as the sparsity rate increases.
This result might be because the Avazu dataset presents more challenges for the methods. Consequently, when averaging the metric results, the overall performance is more heavily influenced by the results from the Avazu dataset.
And PEP, despite being the best for the Criteo dataset, is the worst for Avazu. While TTRec has a more consistent performance across these two datasets. 
\instex{The poor performance of PEP on Avazu compared with that on Criteo can be attributed to differences in dataset sparsity and feature distribution. Though the two datasets have similar numbers of interactions, Avazu has many more features, and the long-tail distribution among features is more severe than Criteo. As PEP uses weight decay to determine the embedding sizes, a larger number of low-frequency features in Avazu are assigned fewer parameters, hurting the final model performance.}

Nonetheless, the gap between the original model and the higher compression rate model in CF is much higher than in CTR prediction (more than 20\% NDCG@20 drop for the most optimal method). There are various possible explanations for this gap. First, the CTR prediction task has multiple fields for a single output, while the CF task only has two fields, which makes it much harder for CF-based models to predict with limited information. Second, most methods are initially tested on the CTR prediction task, making methods provide more competitive results. Third, the CTR model embedding is much bigger than the CF model, thus having more room to compress the model.

\subsection{Real-world Efficiency (RQ3)}




\begin{figure}
    \centering

    \subcaptionbox{Workstation Train Time}
    {
        \vspace{-.2cm}
        \includegraphics[height=0.215\textheight]{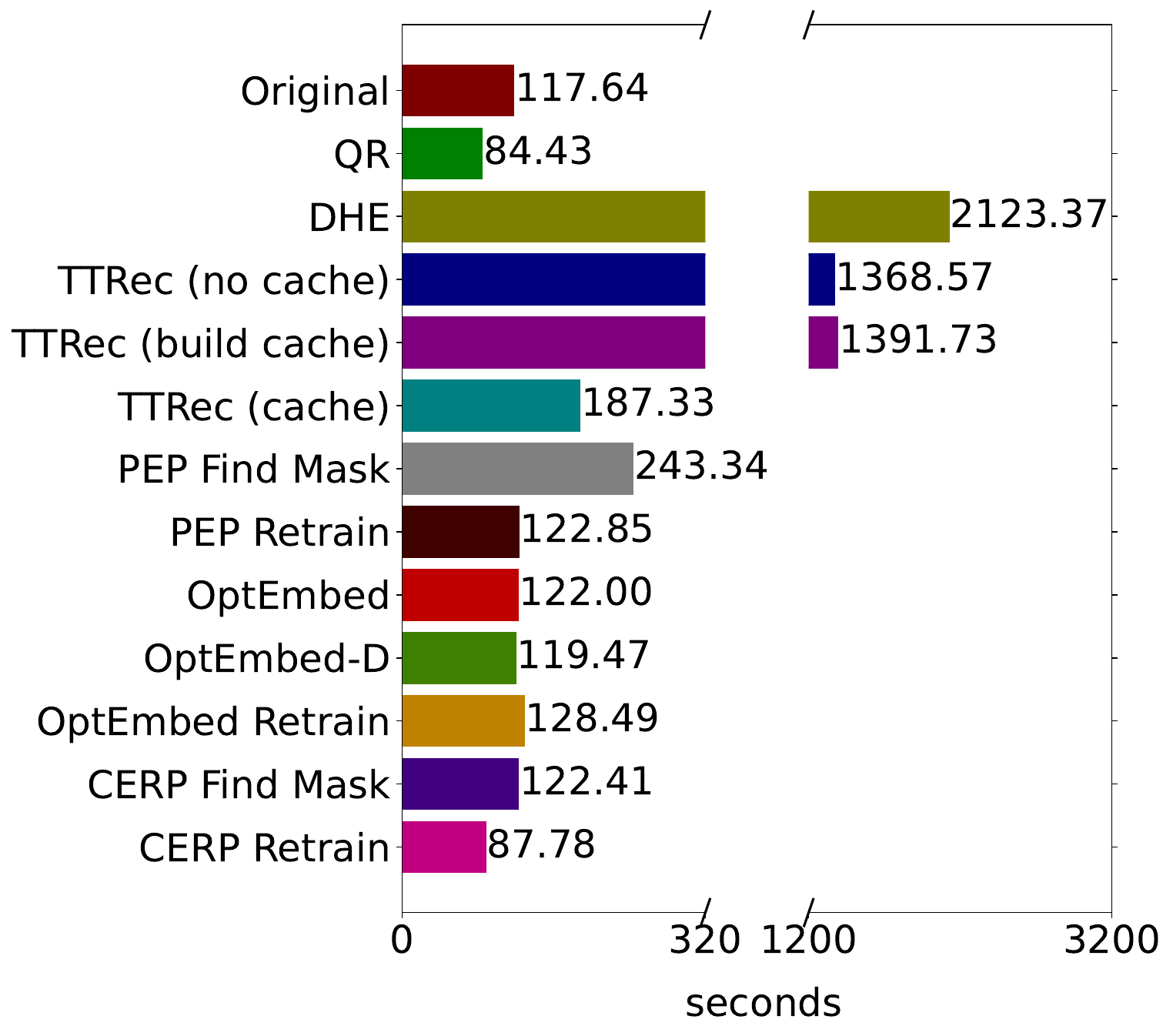}
    }
    \subcaptionbox{Workstation Train VRAM}
    {
        \vspace{-.2cm}
        \includegraphics[height=0.215\textheight]{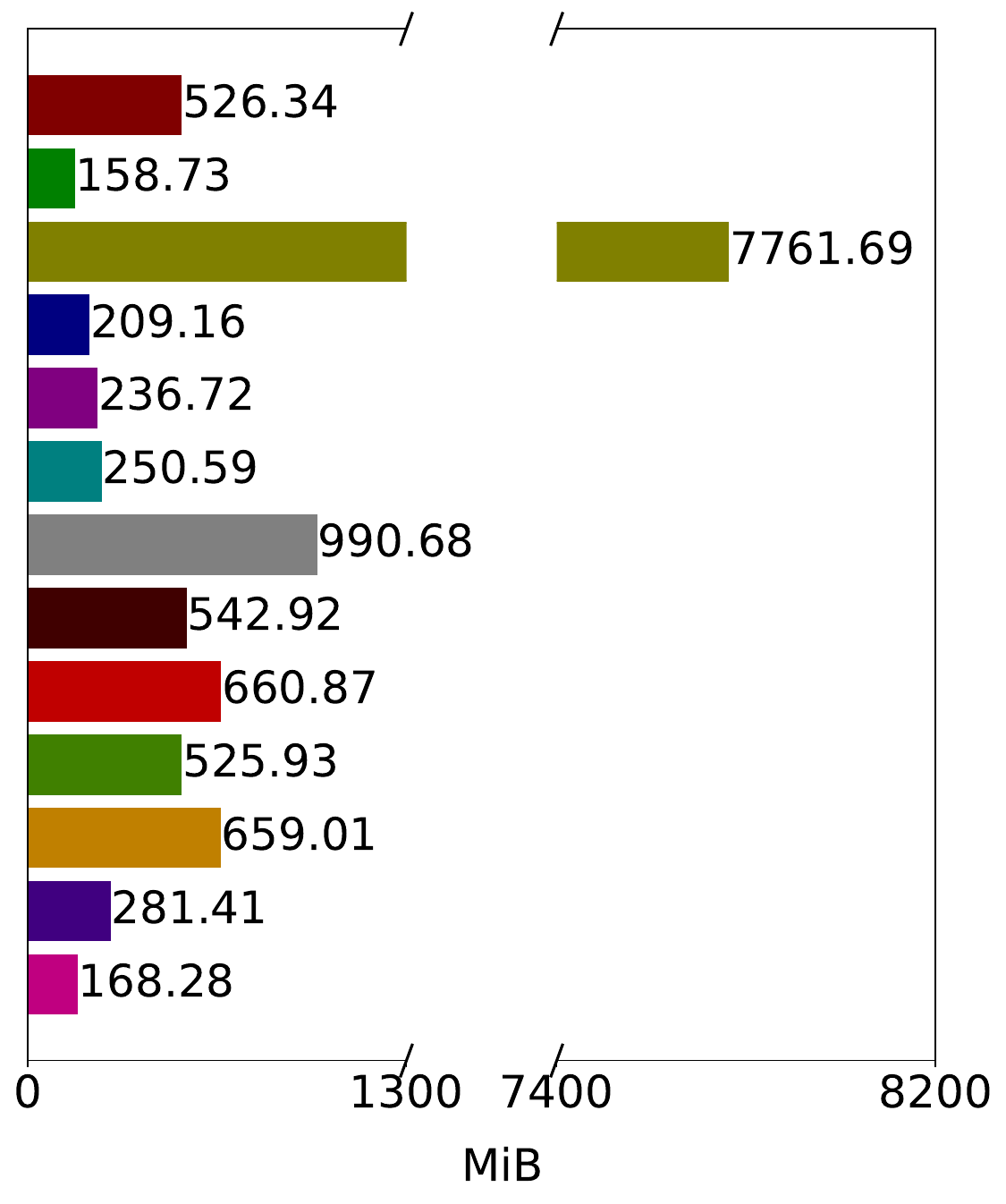}
    }
    \subcaptionbox{Raspberry Train Time}
    {
        \vspace{-.2cm}
        \includegraphics[height=0.215\textheight]{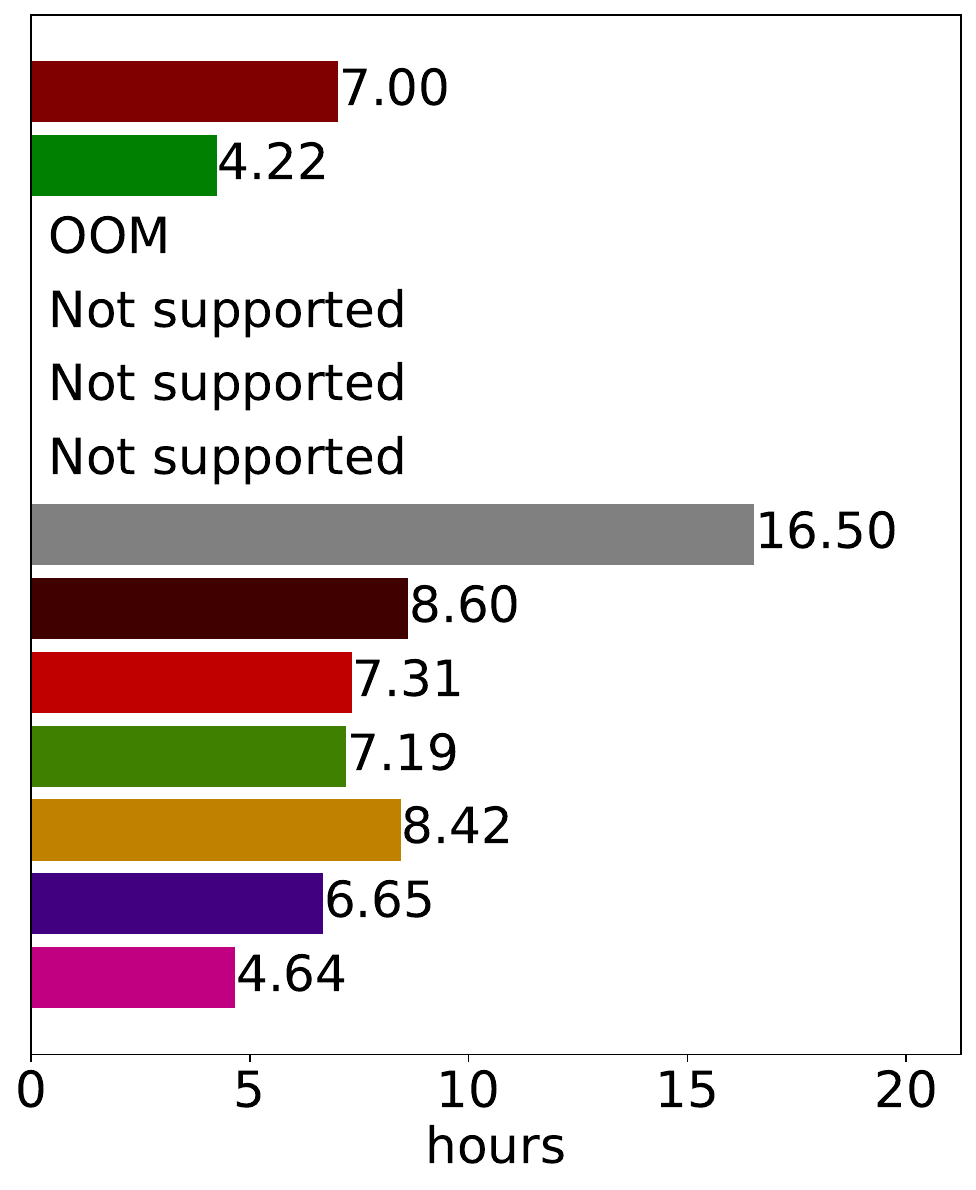}
    }
    \subcaptionbox{Raspberry Train Mem}
    {
        \vspace{-.2cm}
        \includegraphics[height=0.215\textheight]{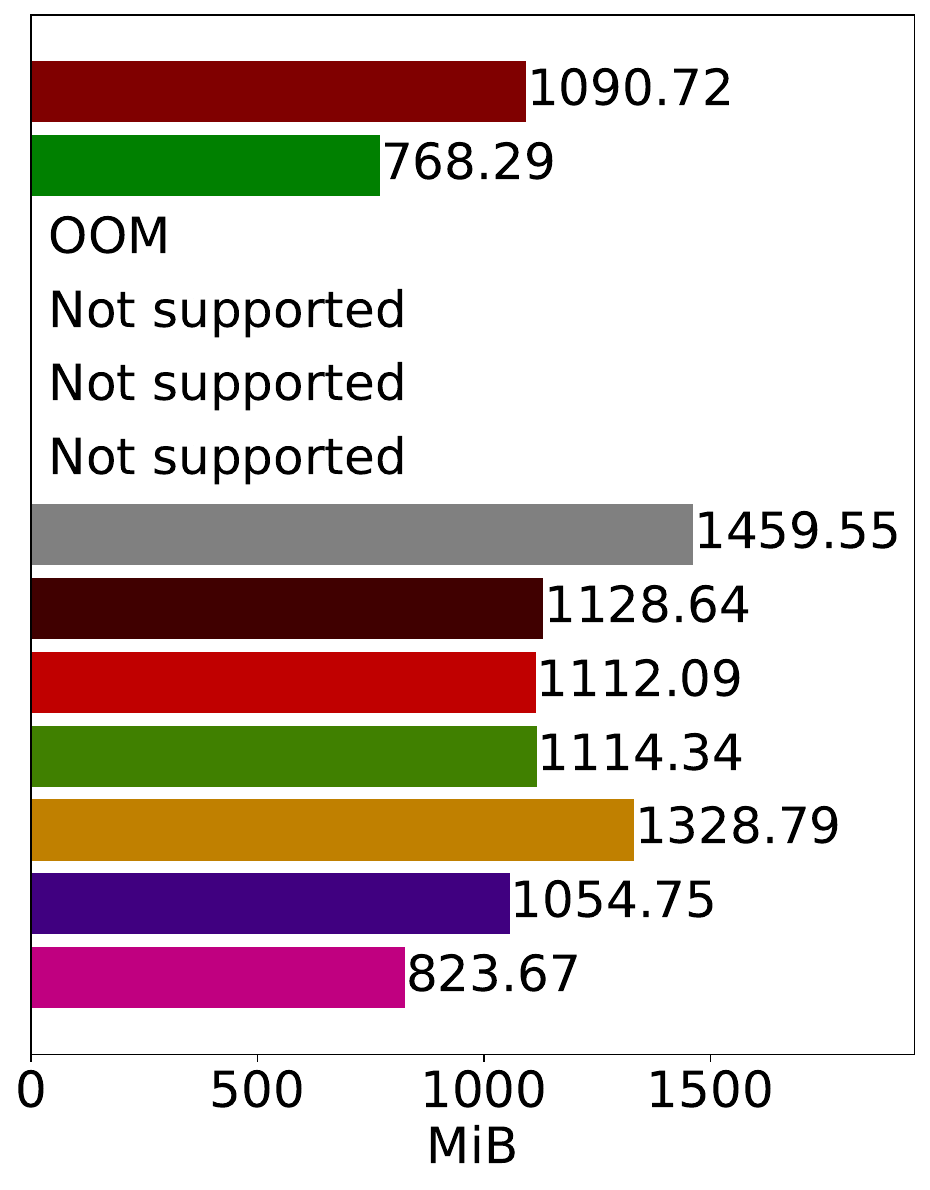}
    }

    \subcaptionbox{Workstation Infer Time}
    {
        \vspace{-.2cm}
        \includegraphics[height=0.16\textheight]{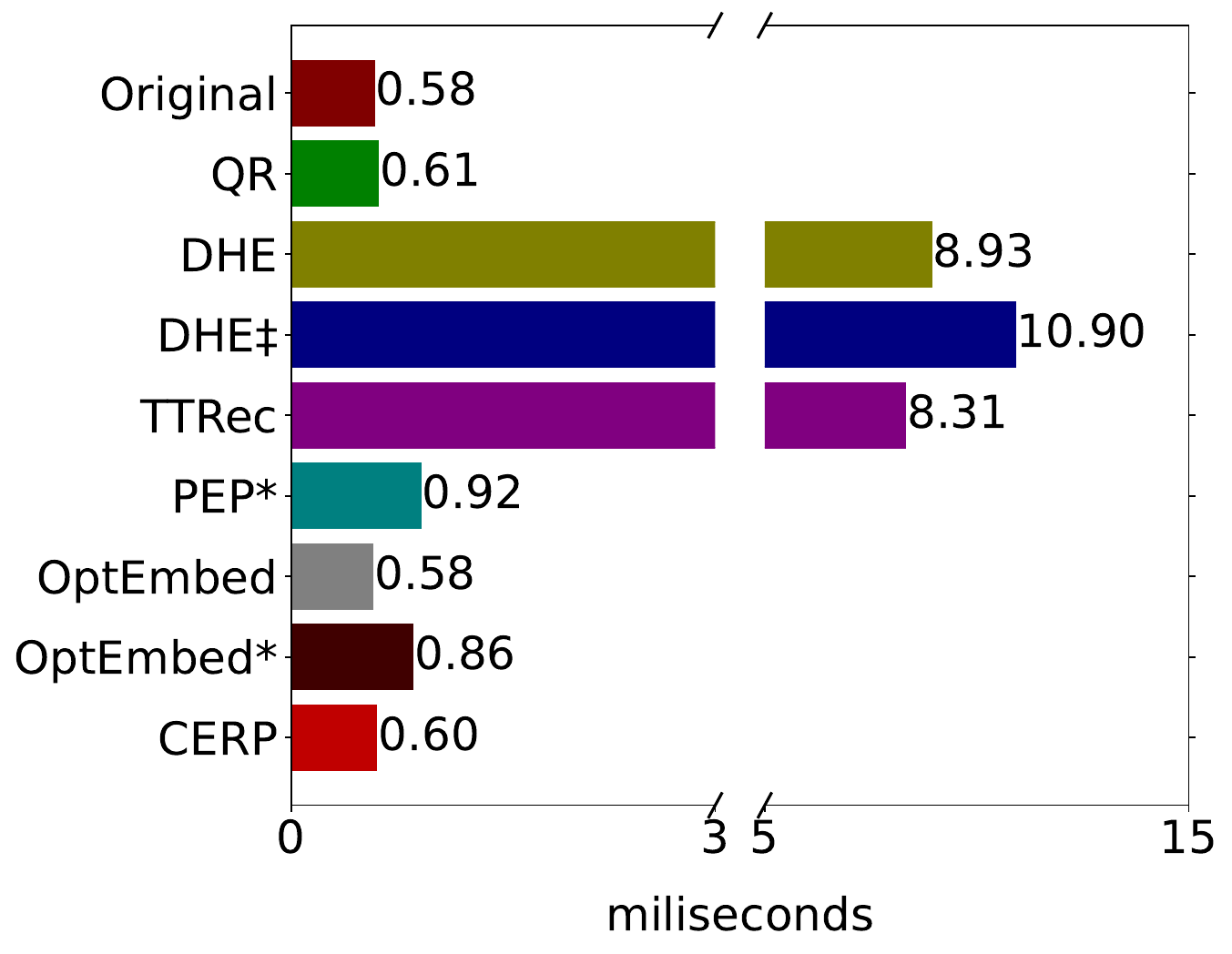}
    }
    \subcaptionbox{Workstation Infer VRAM}
    {
        \vspace{-.2cm}
        \includegraphics[height=0.16\textheight]{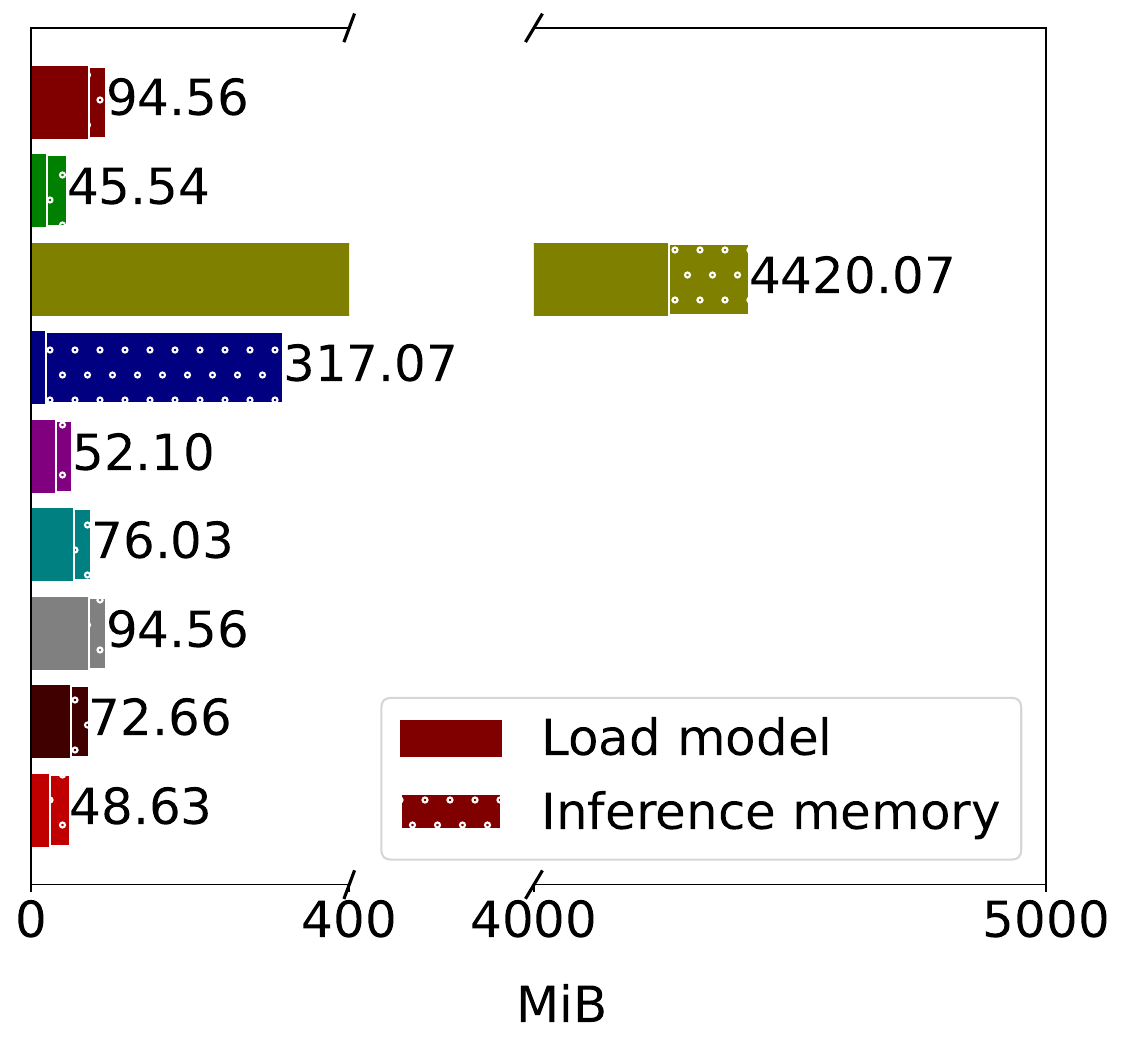}
    }
    \subcaptionbox{Raspberry Infer Time}
    {
        \vspace{-.2cm}
        \includegraphics[height=0.16\textheight]{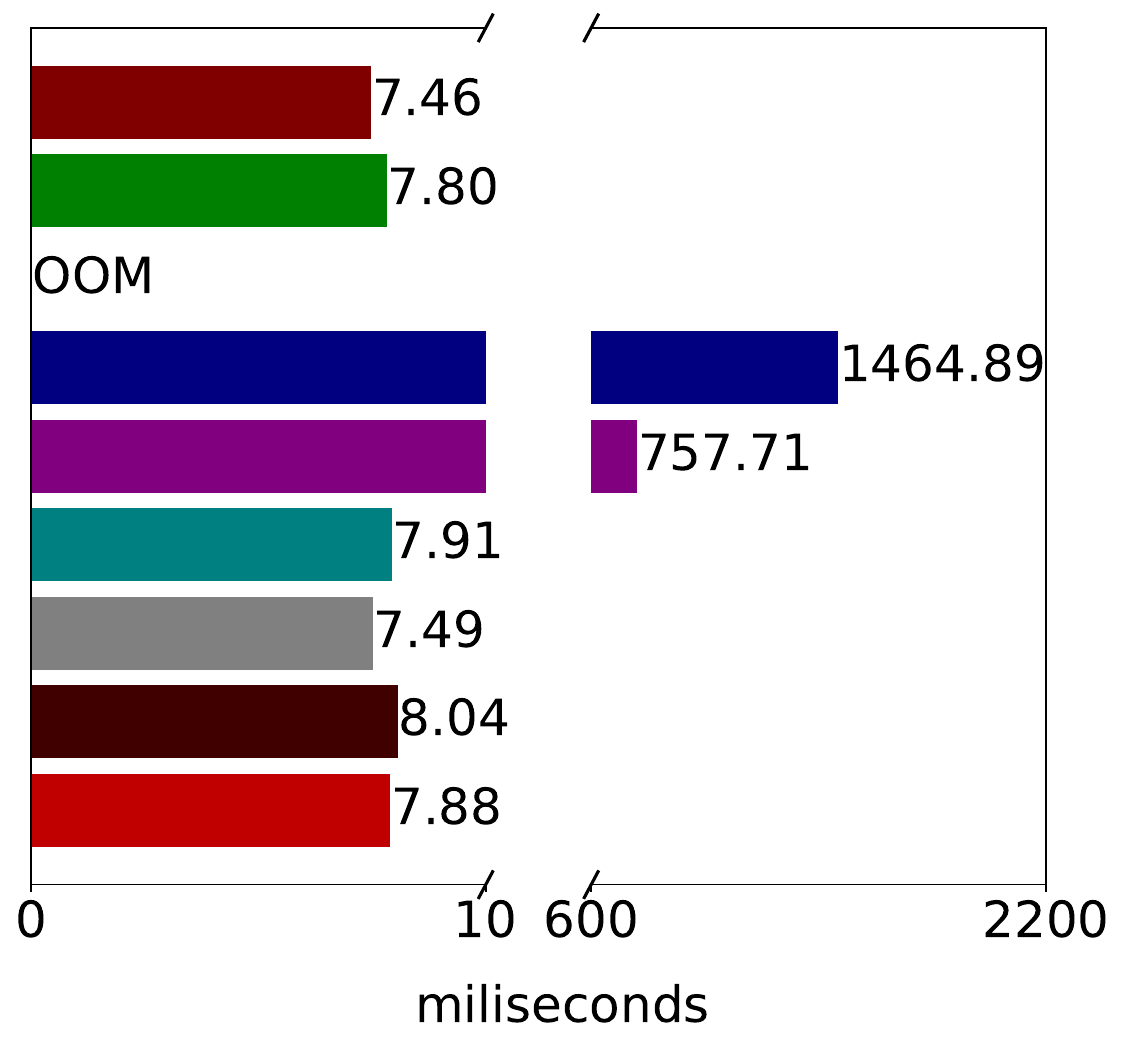}
    }
    \subcaptionbox{Raspberry Infer Mem}
    {
        \vspace{-.2cm}
        \includegraphics[height=0.16\textheight]{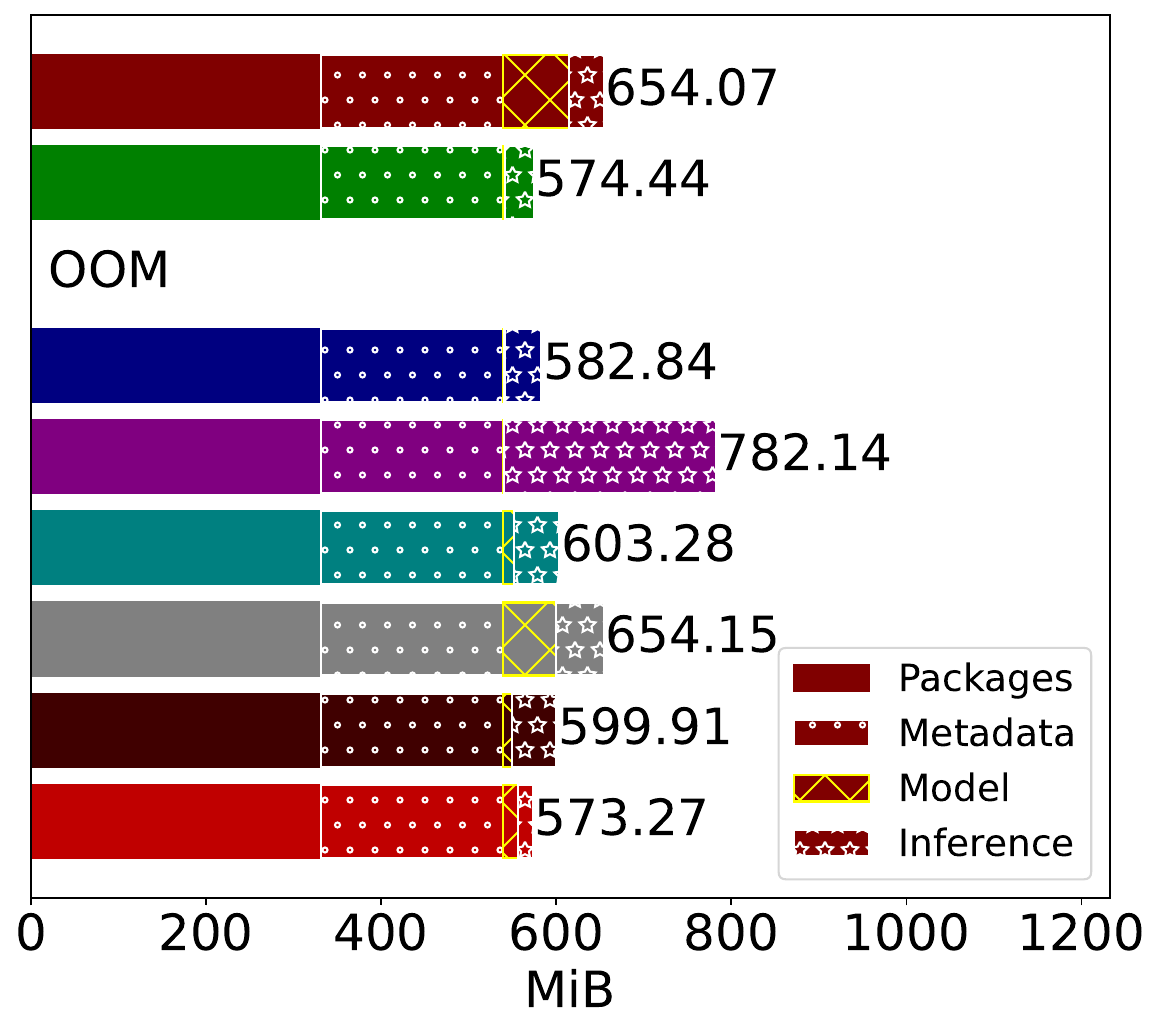}
    }
    \vspace{-.3cm}

    \caption[]%
    {Training and Inference resource usage on the Criteo dataset. The asterisk ``*'' in inference means we store the weight matrix under SparseCSR format. $\ddagger$ means creating hash codes on-the-fly for DHE. TTRec implements a custom CUDA kernel for training, which naturally cannot be implemented on the CPU. TTRec's cache is assumed to be 10\% of the original model. ``Mem'' is a shorthand for memory, ``Packages'' refers to Python packages and overhead in general, and  ``Metadata'' refers to the CPU memory required to store the mapping from features (as string data type) to the corresponding feature IDs (as integer data type).}
    \label{fig:criteo-train-infer-v1}
\end{figure}

\begin{figure}
    \centering
    \subcaptionbox{Workstation Train Time}
    {
        \vspace{-.2cm}
        \includegraphics[height=0.215\textheight]{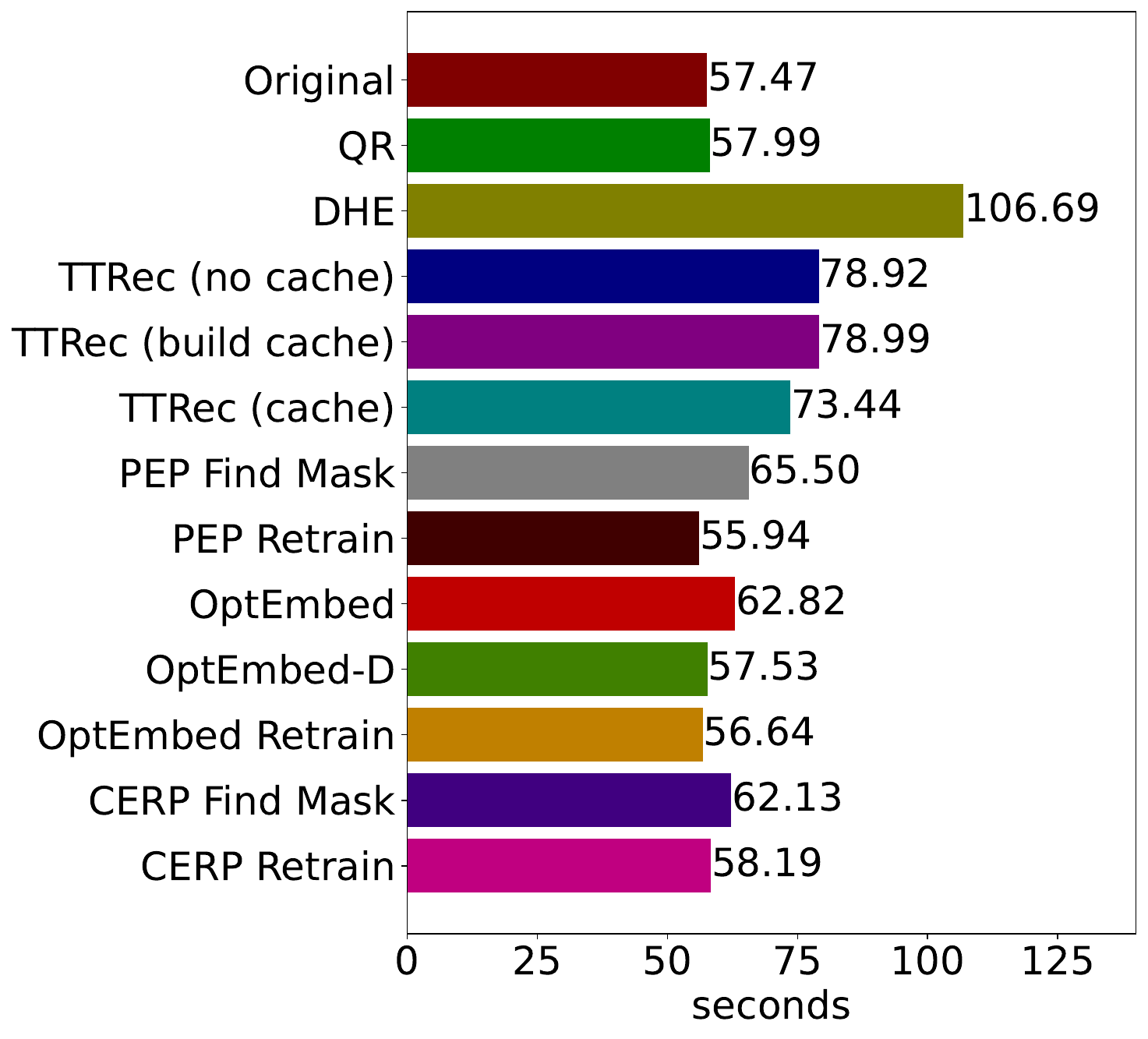}
    }
    \subcaptionbox{Workstation Train VRAM}
    {
        \vspace{-.2cm}
        \includegraphics[height=0.215\textheight]{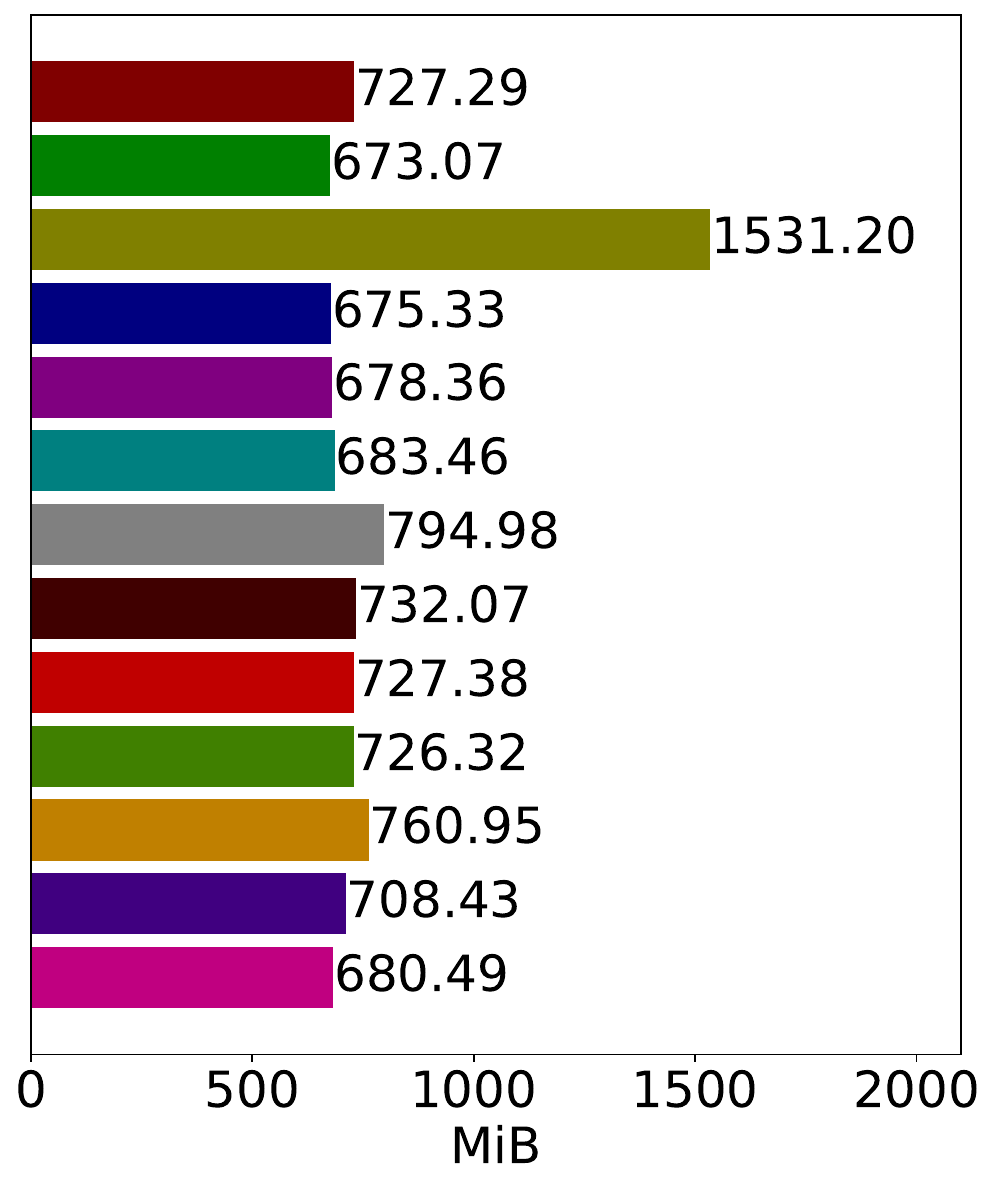}
    }
    \subcaptionbox{Raspberry Train Time}
    {
        \vspace{-.2cm}
        \includegraphics[height=0.215\textheight]{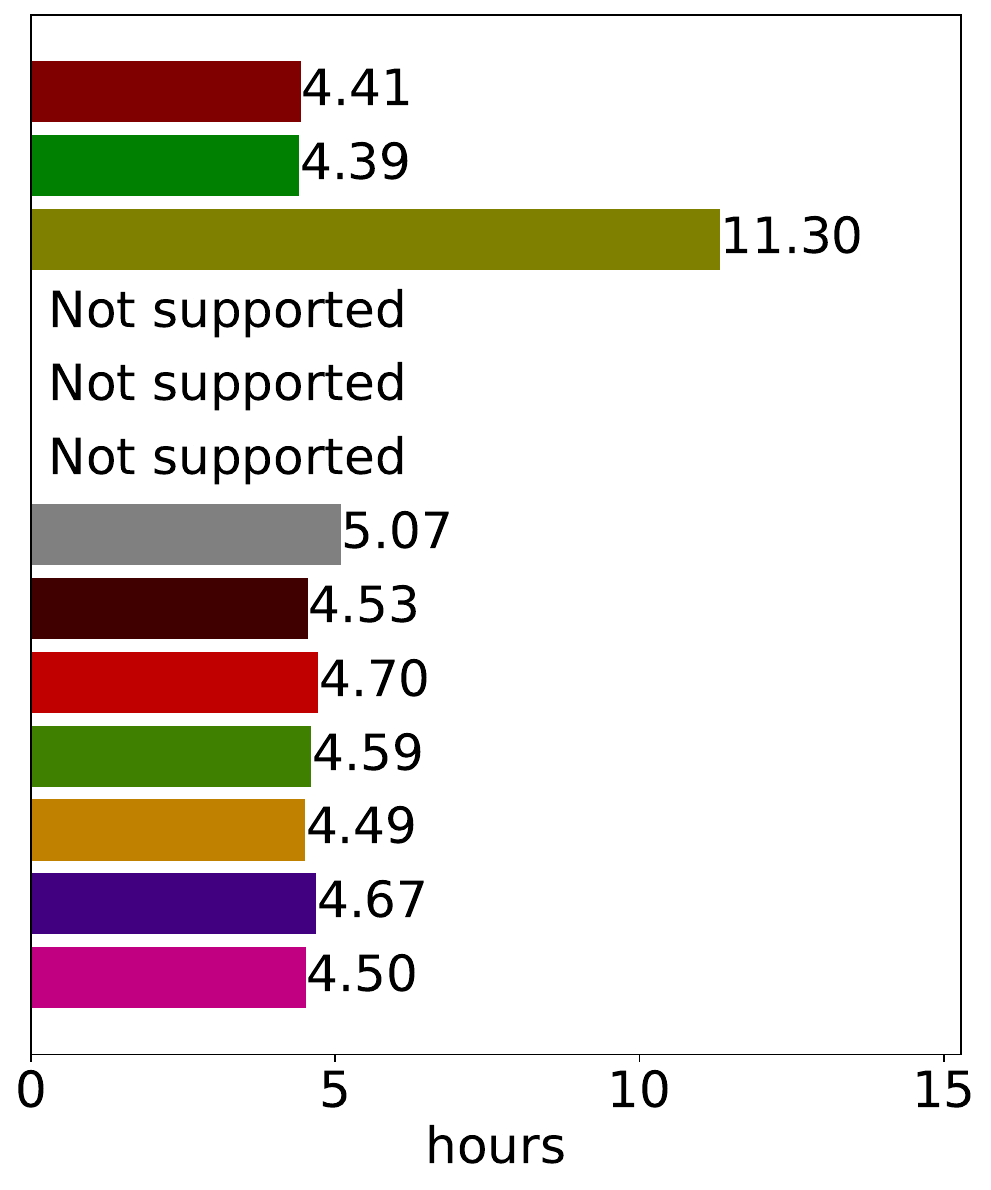}
    }
    \subcaptionbox{Raspberry Train Mem}
    {
        \vspace{-.2cm}
        \includegraphics[height=0.215\textheight]{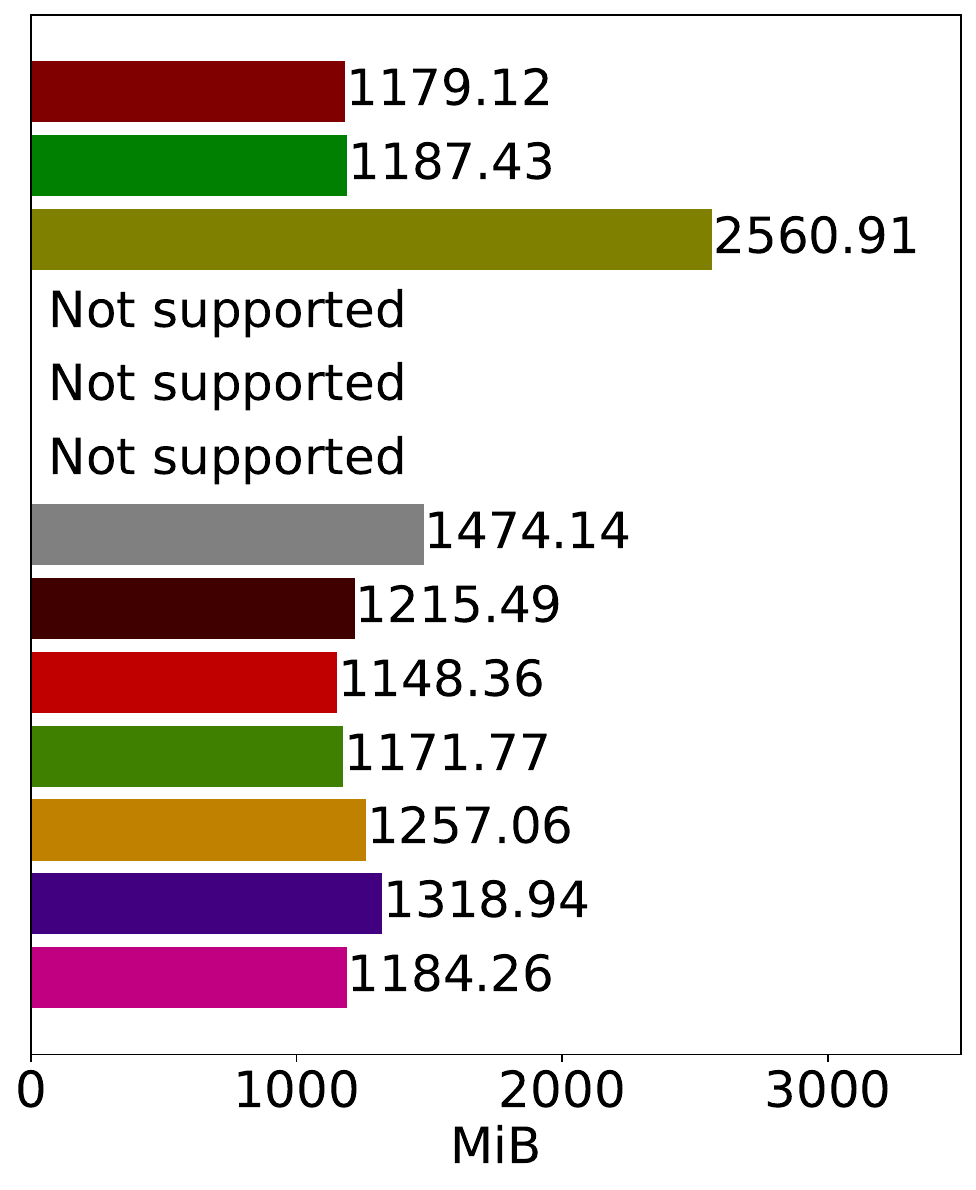}
    }
    
    \subcaptionbox{Workstation Infer Time}
    {
        \vspace{-.2cm}
        \includegraphics[height=0.15\textheight]{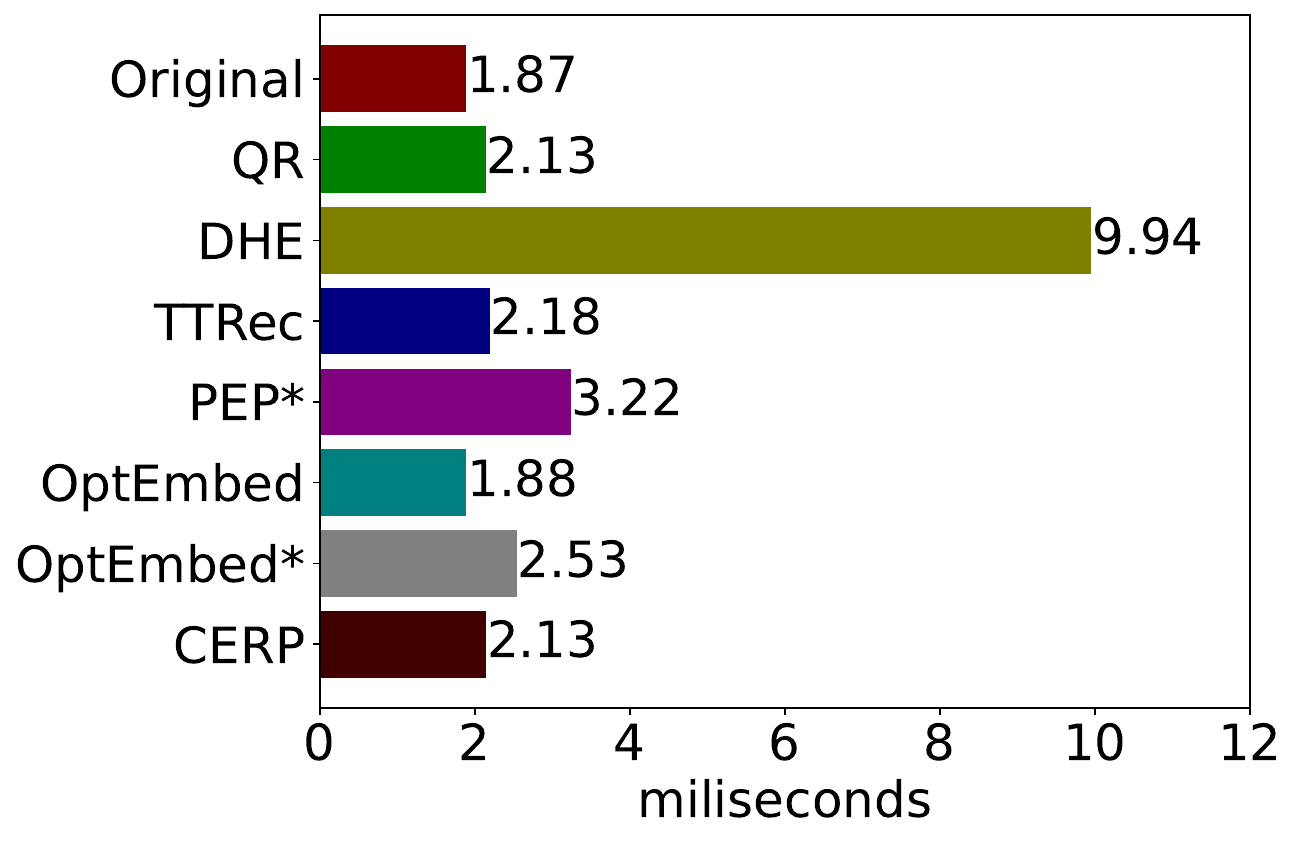}
    }
    \subcaptionbox{Workstation Infer VRAM}
    {
        \vspace{-.2cm}
        \includegraphics[height=0.15\textheight]{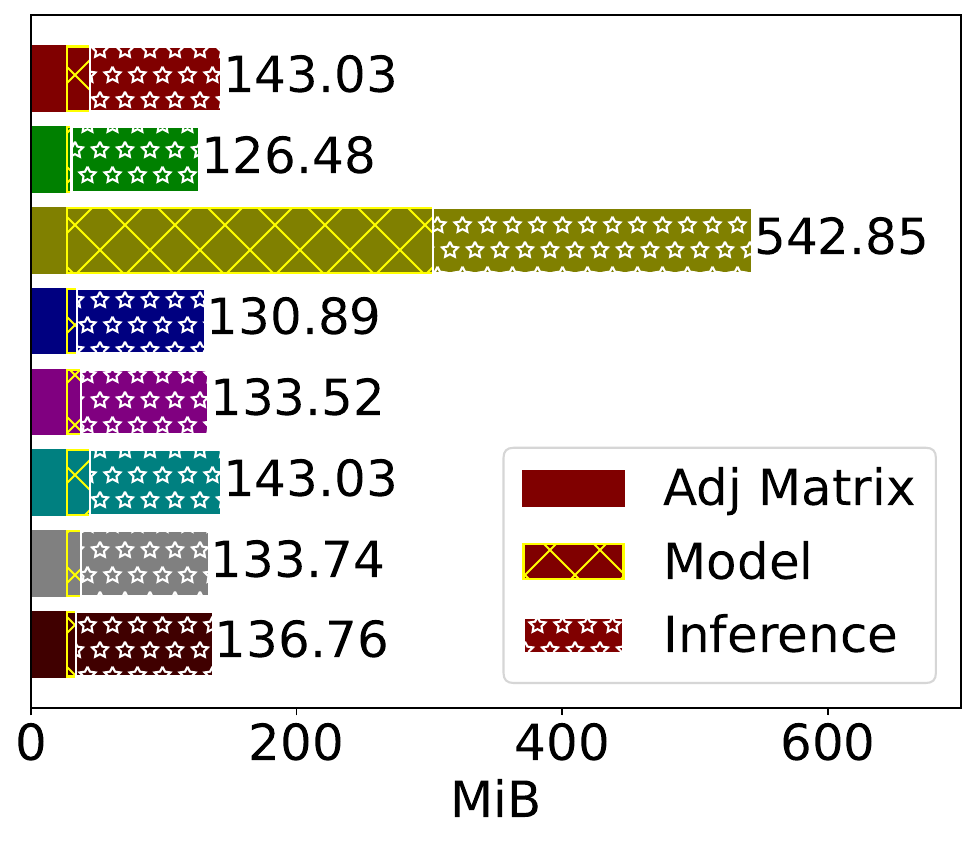}
    }
    \subcaptionbox{Raspberry Infer Time}
    {
        \vspace{-.2cm}
        \includegraphics[height=0.15\textheight]{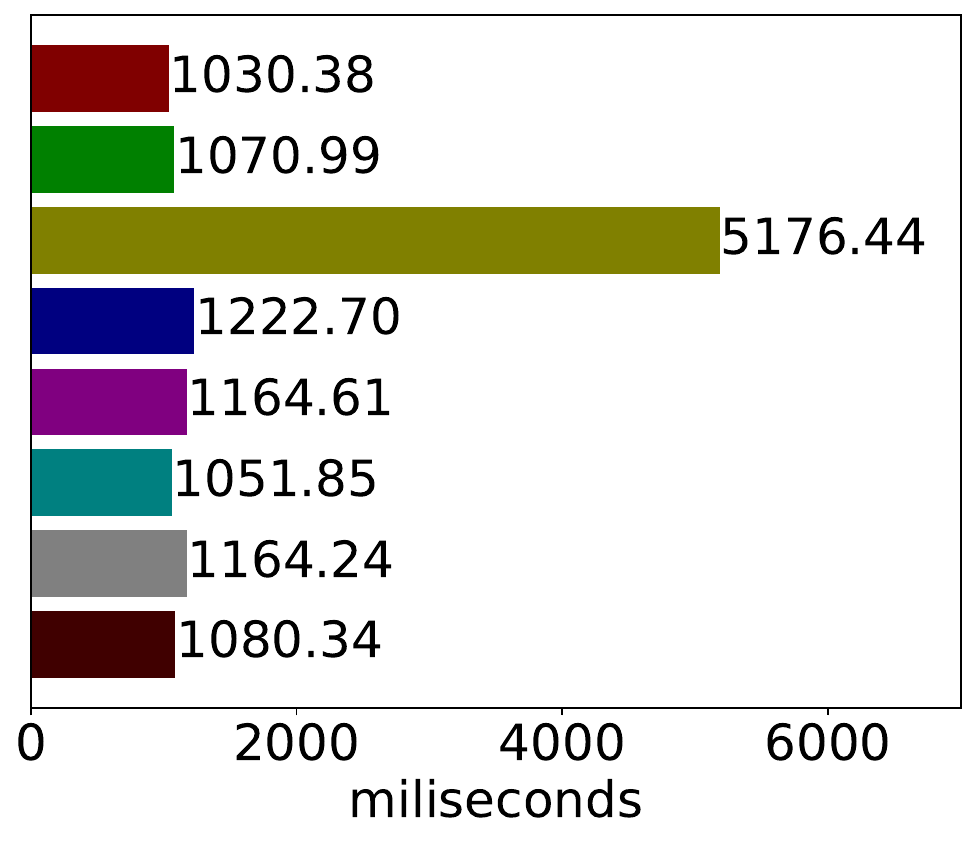}
    }
    \subcaptionbox{Raspberry Infer Mem}
    {
        \vspace{-.2cm}
        \includegraphics[height=0.15\textheight]{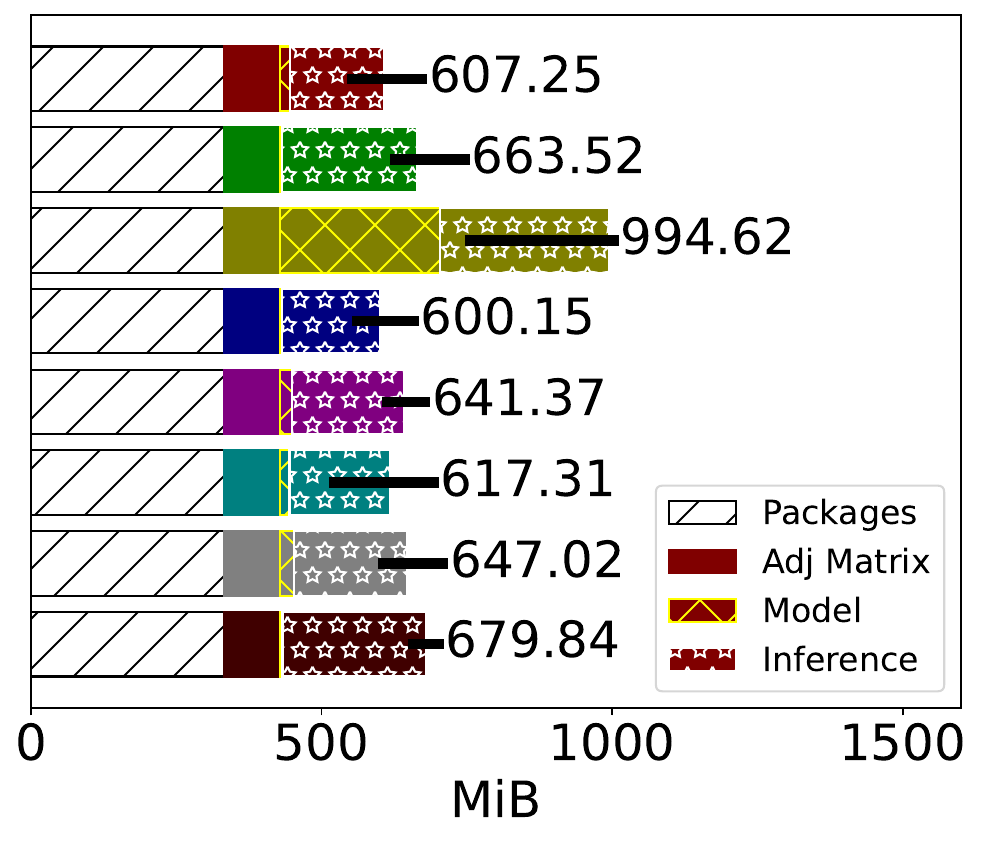}
    }

    \vspace{-.3cm}
    \caption[]%
    {Training and Inference resource usage on Yelp2018 dataset. The asterisk ``*'' in inference means we store the weight matrix under SparseCSR format. 
    TTRec implements a custom CUDA kernel for training, which naturally cannot be implemented on the CPU. TTRec's cache is assumed to be 10\% of the original model. ``Mem'' is a shorthand for memory, ``Packages'' refers to Python packages and overhead in general, and  ``Metadata'' refers to the CPU memory required to store the mapping from features (as string data type) to the corresponding feature IDs (as integer data type).}
    \label{fig:yelp-eff}
\end{figure}

Figure \ref{fig:criteo-train-infer-v1} shows the resource usage in training and inference in the Criteo dataset, and figure \ref{fig:yelp-eff}  depicts the resource usage for the Yelp2018 dataset. 
\instex{The packages are Python library data required to be loaded on RAM (mainly PyTorch), whose memory consumption is independent from the models used. Package memory consumption can be reduced by using more compact coding languages such as C++. Metadata is mainly the hash dictionary that maps the raw string of features to its corresponding ID, which is especially memory-intensive as it increases with the number of features. There are techniques to reduce the memory cost of this dictionary \cite{tim2021}, but they are outside the scope of this work as we lay emphasis on the memory consumption from the model itself. 
It is worth noting that the Raspberry Pi environment is different from the workstation setting. This is because packages and metadata do not exist in VRAM used by workstation's GPU, and only exist in the CPU RAM of Raspberry Pi. } 
As some methods are comprised of multiple steps, we benchmark those intermediate steps separately for a clearer view of each method's performance:
\begin{itemize}
    \item TTRec \cite{ttrec2021}: We test two main scenarios for TTRec training, i.e., with and without cache. The original paper suggests storing the most accessed embedding vectors in an uncompressed format, requiring a few warm-up iterations (e.g., the first epoch) to build this cache. This step is denoted as ``build cache''. Once the cache is built, it is treated as normal trainable parameters and directly updated with backward propagation in the following iterations besides the TT-cores. To study the overhead introduced for the cache-building process, we include the without-cache configuration in our experiment. 
    \item ``Find Mask'' and ``Retrain'': Most pruning methods first find the embedding mask, followed by a retraining step. If this is the case, we provide results for each step separately.
    \item OptEmbed \cite{optembed2022}: We benchmark two ``Find Mask'' versions for OptEmbed: OptEmbed (original model) and OptEmbed-D (removes feature mask $m_e$).
\end{itemize}

We could train all methods on Raspberry Pi except for TTRec, which is designed specifically for GPU.
Among the methods, QR, the simplest, uses the least resource, while DHE is the most resource-consuming as it requires an MLP model inside. 
Overall, an increased memory footprint is mostly observed during training, and the majority of the methods can improve their memory usage during inference. As for the additional training memory overhead, pruning-based methods (PEP, OptEmbed) yield increased memory usage due to the embedding mask used during fine-tuning, and the full embedding table has to be in place for learning the embedding mask as well. For DHE, the overhead stems from computing high-dimensional hash codes and training an MLP to generate dense embeddings on-the-fly with those codes. For instance, the hash code dimension (1024) is much larger than the embedding dimension (16) in DHE, leading to a significant memory overhead. In contrast, methods using compositional embeddings like QR, TTRec and CERP can effectively cut the memory usage in most cases because of the inherently smaller parameter space they use to represent all features.
It is worth noting that DHE performance could be further improved with other methods from lightweight machine learning models such as pruning and quantization; however, this is outside of our research scope. 
Although training on edge devices is theoretically possible, it remains impractical in most scenarios, as training a single epoch requires over four hours due to the limited computational resources available.

In the context of the CTR task, TTRec demands significant resources during its first epoch to create a cache but demonstrates substantially increased efficiency in the subsequent phase. 
Compositional-based methods (except for DHE) and CERP have better training efficiency than the original model in both VRAM and time consumption.
In contrast, most pruning-based methods use more resources during the training phase as they require extra memory to store the found mask. Another key contributor to the training efficiency is the $L_2$ regularization optimization. While the gradient for log loss is sparse, the gradient for $L_2$ regularization is dense. Given the large amount of RSs' parameters, this results in slow computation when updating the model's parameters. Compositional-based methods benefit from this as they have less parameter count in the training phase. In contrast, pruning methods still retain their dense format in the training phase.

Regarding top-$k$ recommendation efficiency, the memory inference cost is much higher than the model size.
There are two main reasons for this. First, LightGCN requires extracting the full embedding tables for feedforward. 
Second, the intermediate layer results dominate memory consumption.
So, despite saving the storage cost, smaller LightGCN models don't benefit from saved memory costs. 
Consequently, training and inference memory usage differ much less significantly between each method, and no method could reduce both time and memory consumption in training efficiency. 
Similar trends with the CTR task, such as pruning consuming more training resources, are also observed.

\subsubsection{Overall Summary of The Real-world Efficiency Test.} \instex{It is worth mentioning that, none of our chosen methods outperforms the original model regarding inference runtime. This is because the compressed model parameters generally incur some computational overhead when the model is being decompressed during inference, e.g., TTRec requires a sequence of tensor multiplication operations to recover the full-size embedding table.
However, methods that use less memory could benefit from larger batch sizes, and LERSs could also reduce inference time when required to transfer embedding from bulk storage (e.g., disk) to faster storage (e.g., RAM, VRAM) \cite{hsia2023mp}, but this is outside our research scope.
Compositional embedding methods creates embeddings from meta vectors, while pruning pays the extra overhead in accessing embeddings stored in a sparse matrix format.}


\begin{table}[t!]
    \centering
    \begin{tabular}{lrrrrrrrr}
    \toprule
        \multirow[b]{3}{*}{Method} &  \multicolumn{4}{c}{CTR}
        & \multicolumn{4}{c}{CF} \\
    \cmidrule(lr){2-5}
    \cmidrule(lr){6-9}
    & \multicolumn{2}{c}{Runtime} & \multicolumn{2}{c}{Memory} & \multicolumn{2}{c}{Runtime} & \multicolumn{2}{c}{Memory} \\
    \cmidrule(lr){2-3}
    \cmidrule(lr){4-5}
    \cmidrule(lr){6-7}
    \cmidrule(lr){8-9}
    & Abs. & Relative & Abs. & Relative & Abs. & Relative & Abs. & Relative \\  
    \midrule
QR        & 0.03  & 5.17\%    & -49.02 & -51.84\% & 0.26 & 44.83\%   & -16.55 & -17.50\% \\
DHE       & 10.32 & 1779.31\% & 222.51 & 235.31\% & 8.07 & 1391.38\% & 399.82 & 422.82\% \\
TTRec     & 7.73  & 1332.76\% & -42.46 & -44.90\% & 0.31 & 53.45\%   & -12.14 & -12.84\% \\
PEP      & 0.34  & 58.62\%   & -18.53 & -19.60\% & 1.35 & 232.76\%  & -9.51  & -10.06\% \\
OptEmbed & 0.28  & 48.28\%   & -21.9  & -23.16\% & 0.66 & 113.79\%  & -9.29  & -9.82\%  \\
CERP      & 0.02  & 3.45\%    & -45.93 & -48.57\% & 0.26 & 44.83\%   & -6.27  & -6.63\%  \\
    \bottomrule
    \end{tabular}
    
    \caption{Absolute (Abs.) and relative differences in efficiency metrics of various methods during inference compared to original models in the workstation settings.}
    \label{tab:tradeoff}
\end{table}

\subsubsection{Additional Results on The Computational Overhead from Embedding Compression}
\instex{To better illustrate the overhead differences, we calculate the runtime and memory differences between the original model and their compressed versions, and the results are shown in
Table \ref{tab:tradeoff}. 
Firstly, QR provides the smallest sacrifice in efficiency, evidenced by the lowest runtime overhead and largest memory reduction. DHE has the highest runtime overhead and does not reduce peak memory consumption during inference. Secondly, DHE  is able to reduce the parameter consumption from embedding table, in our case, DHE's heavy MLP component has offset the embedding parameter reduction.
Lastly, despite both PEP and OptEmbed are pruning-based methods and use the same underlying implementation with SparseCSR \cite{hoefler2021sparsity}, OptEmbed is faster than PEP due to a more hardware-friendly implementation.}

\begin{figure}[t]
    \centering
    \subcaptionbox{Yelp2018}
    {
        \includegraphics[width=0.48\textwidth]{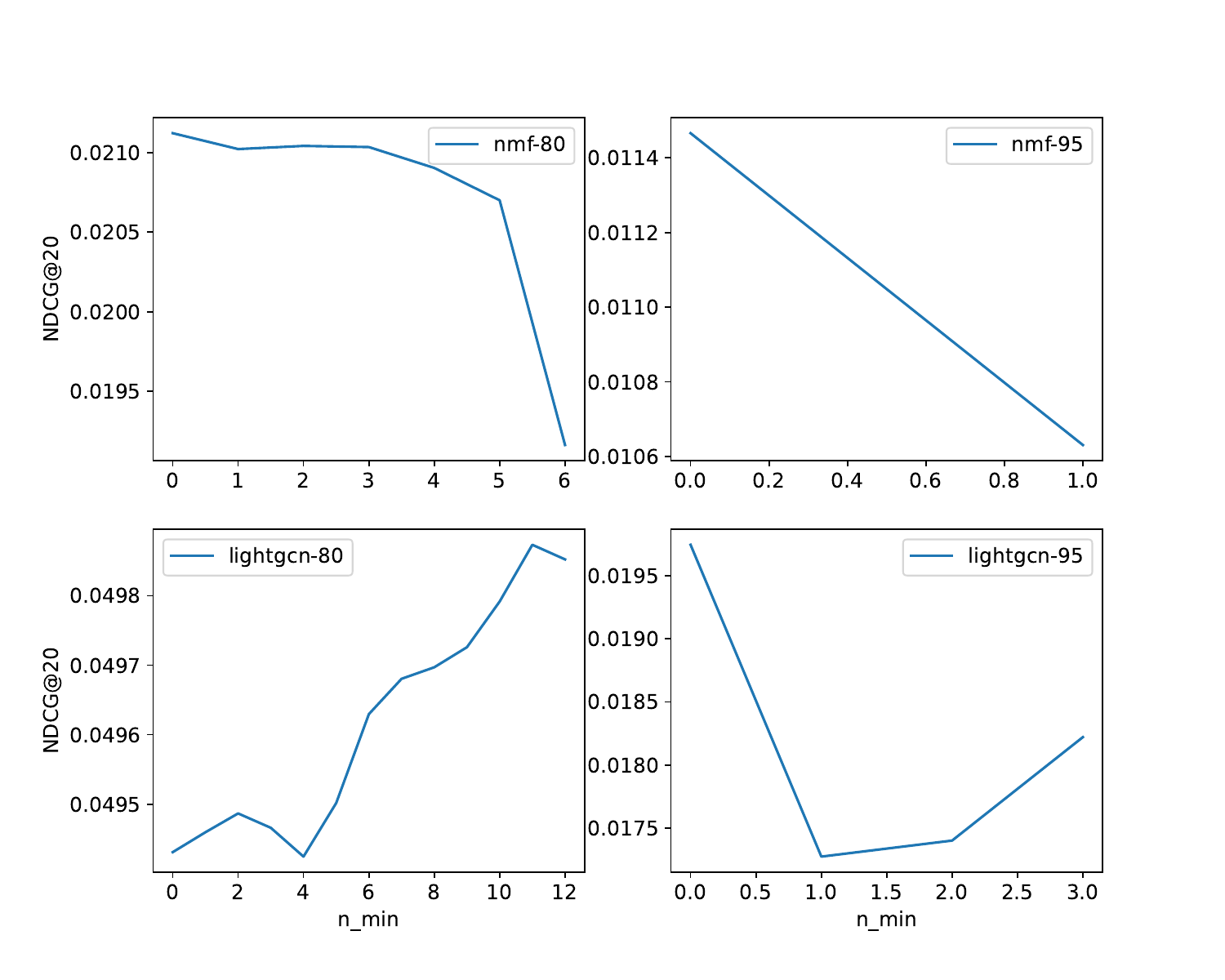}
    }
    \subcaptionbox{Gowalla}
    {
        \includegraphics[width=0.48\textwidth]{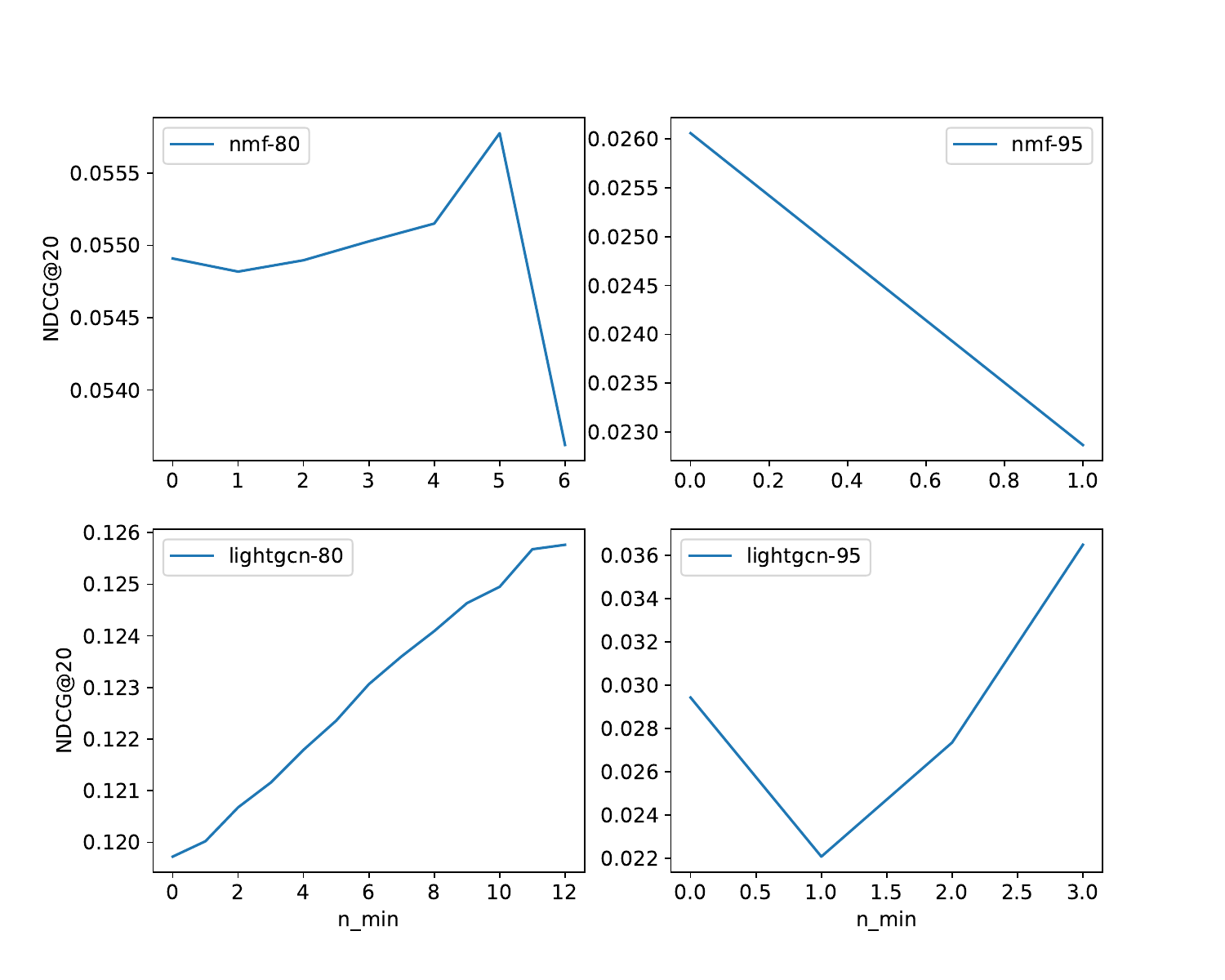}
    }
    \caption[]%
    {Performance of MagPrune with different $n\_min$. For Yelp2018 and Avazu, the backbones are respectively LightGCN and DCN.}
    \label{fig:mag-prune}
\end{figure}

\subsection{Ablation Study on MagPrune}

In this section, we explore the effect of $n\_min$ in MagPrune performance. Figure \ref{fig:mag-prune} shows the performance of MagPrune across different $n\_min$. 
In general, optimizing $n\_min$ can further improve model performance, especially under moderate sparsity. 
These results suggest that ensuring each user and item has at least some non-zero value can positively impact the performance of LightGCN. 
However, it is crucial to determine $n\_min$ through systematic optimization rather than arbitrary selection, as too high $n\_min$ can decrease performance.

\subsection{\instex{Study on Auxiliary Loss Functions}}

\begin{figure}[ht]
    \centering
    \pgfplotstableread[col sep=comma]{data/yelp_wd.csv}\yelpwd
\pgfplotstableread[col sep=comma]{data/avazu_wd.csv}\avazuwd

\begin{center}
    \begin{tikzpicture}    
    \begin{axis}[
        title={\large Yelp2018},
        xmin=-0.7, xmax=2.7,
        xtick={0, 1, 2},
        xticklabels={50\%, 80\%, 95\%},
        ybar,
        bar width=12pt,
        width=0.47\textwidth,
        xtick=data,
        ymin=0, 
        ylabel=Weight Decay,
        xlabel=Sparsity Rate,
        legend style={
            anchor=north,
            legend columns=-1,
        },
        legend to name=sharedlegend
    ]
    \addplot table[
        x expr=\coordindex, 
        y=qr
    ]
    {\yelpwd};
    \addplot table[
        x expr=\coordindex,
        y=dhe
    ]
    {\yelpwd};

    \addplot table[
        x expr=\coordindex,
        y=pep
    ]
    {\yelpwd};
    
    \addplot[
        red, dashed,
        line legend,
        sharp plot,
        update limits=false
    ] coordinates {(-1,0.00975999679147741) (5,0.00975999679147741)};
    
    \legend{QR;, DHE;, PEP;, Original}
    \end{axis}
    \end{tikzpicture}
    \begin{tikzpicture}    
    \begin{axis}[
        title={\large Avazu},
        xmin=-0.7, xmax=2.7,
        xtick={0, 1, 2},
        xticklabels={50\%, 80\%, 95\%},
        ybar,
        bar width=12pt,
        width=0.47\textwidth,
        xtick=data,
        ymin=0, ymax=0.0035,
        xlabel=Sparsity Rate,
        legend style={
            at={(0.8, 1.0)},
            anchor=north,
        },
    ]
    \addplot table[
        x expr=\coordindex, 
        y=qr
    ]
    {\avazuwd};
    \addplot table[
        x expr=\coordindex,
        y=dhe
    ]
    {\avazuwd};

    \addplot table[
        x expr=\coordindex,
        y=pep
    ]
    {\avazuwd};

    \addplot[
        red, dashed,
        line legend,
        sharp plot,
    ] coordinates {(-1,0.0033289440083949075) (5,0.0033289440083949075)};
    
    \end{axis}
    \end{tikzpicture}
    {       
    \begin{tikzpicture}
        \node at (0,0) {
            \pgfplotslegendfromname{sharedlegend}
        };
    \end{tikzpicture}
    }
\end{center}
    \vspace{-.5cm}
    \caption{The relationship between the weight decay and the sparsity rate. For Yelp2018 and Avazu, the backbones are respectively LightGCN and DCN.}
    \label{fig:wd}
\end{figure}

\begin{figure}[ht]
    \centering
    \pgfplotstableread[col sep=comma]{data/yelp_infonce.csv}\loadtable

\begin{center}
    \begin{tikzpicture}
    \begin{axis}[
        title={\large Yelp},
        xmin=-0.7, xmax=2.7,
        xtick={0, 1, 2},
        xticklabels={50\%, 80\%, 95\%},
        ybar,
        bar width=12pt,
        width=0.47\textwidth,
        xtick=data,
        ymin=0, 
        ylabel=InfoNCE Coefficient,
        xlabel=Sparsity Rate,
            legend style={
            at={(0.8, 1.0)},
            anchor=north,
            legend columns=-1,
        },
        legend to name=sharedlegend
    ]
    \addplot table[
        x expr=\coordindex, 
        y=qr
    ]
    {\loadtable};
    \addplot table[
        x expr=\coordindex,
        y=dhe
    ]
    {\loadtable};

    \addplot table[
        x expr=\coordindex,
        y=pep
    ]
    {\loadtable};

    \addplot[
        red, dashed,
        line legend,
        sharp plot
    ] coordinates {(-1, 0.6) (5, 0.6)};
    
    \legend{QR;, DHE;, PEP;, Original}
    \end{axis}
    \end{tikzpicture}
    \pgfplotstableread[col sep=comma]{data/gowalla_infonce.csv}\loadtable
    \begin{tikzpicture}
    \begin{axis}[
        title={\large Gowalla},
        xmin=-0.7, xmax=2.7,
        xtick={0, 1, 2},
        xticklabels={50\%, 80\%, 95\%},
        ybar,
        bar width=12pt,
        width=0.47\textwidth,
        xtick=data,
        ymin=0, 
        xlabel=Sparsity Rate,
    ]
    \addplot table[
        x expr=\coordindex, 
        y=qr
    ]
    {\loadtable};
    \addplot table[
        x expr=\coordindex,
        y=dhe
    ]
    {\loadtable};

    \addplot table[
        x expr=\coordindex,
        y=pep
    ]
    {\loadtable};

    \addplot[
        red, dashed,
        line legend,
        sharp plot
    ] coordinates {(-1, 0.9) (5, 0.9)};
    
    \end{axis}
    \end{tikzpicture}

    \begin{tikzpicture}
        \node at (0,0) {
            \pgfplotslegendfromname{sharedlegend}
        };
    \end{tikzpicture}
\end{center}
    \vspace{-.5cm}
    \caption{The relationship between the InfoNCE coefficient and sparsity rates.}
    \label{fig:infonce}
\end{figure}

\instex{As a common practice for recommendation model training, additional optimization objectives like $L_2$ regularization and self-supervision loss (e.g., InfoNCE) can potentially improve the generalizability of the trained model. By default, we have also applied both $L_2$ and InfoNCE losses on top of the recommendation objective. In this section, we aim to further study the potential impact from those additional loss functions. Specifically, we showcase the best loss coefficient settings for three methods, namely QR, DHE, and PEP for demonstration.} 

\instex{Figure \ref{fig:wd} shows the relationship between weight decay ($L_2$ coefficient) and sparsity rate. 
We observe that higher sparsity model prefers lower weight decay coefficient. Since these models are smaller, they are less prone to overfitting, lowering the demand for strong weight decays. Figure \ref{fig:infonce} depicts the relationship between InfoNCE and sparsity rates. 
Interestingly, we observe a similar trend with weight decay. One possible hypothesis is that InfoNCE improves LightGCN's performance by reducing the dimensional collapse (``over-squashing'') phenomenon \cite{simgcl2022,chen2024towards}, where the embeddings only span a low-dimension subspace instead of the entire feature space. When the model becomes smaller/sparse, there would be less dimensional collapse, reducing the necessity of InfoNCE loss.}


\section{Conclusions and Recommendations}
In this study, we conducted a comprehensive benchmark of various embedding compression methods on two main tasks of recommendation models. Our evaluation highlighted the trade-offs between performance and efficiency, offering valuable insights for both researchers and practitioners. However, exciting questions remain for future researchers to explore: 
\begin{itemize}
\item \textbf{Suggestions for future research on LERSs:} First, since most methods currently hinder training efficiency, especially for graph-based collaborative filtering, we suggest that further studies aim to improve this aspect. Second, we recommend incorporating magnitude pruning as a simple yet effective baseline for future research on LERSs. Third, we encourage future researchers to include benchmarks for training and inference efficiency to better demonstrate their methods' effectiveness, particularly for inference, which is often overlooked or unimplemented.

\item \textbf{Relationship between dataset, compression rate, and method performance:} Our experiments reveal varying performances across recommendation datasets; however, the reasons for these disparities remain unclear. Zhang et al. \cite{zhang2023experimental} observed similar patterns in their CTR task experiments. This opens a critical question for future research on the relationship between dataset characteristics and method performance.

\item \textbf{Further evaluations:} While our study provides valuable insights, various metrics, such as energy consumption, diversity, and novelty, remain unexplored despite their importance for edge devices \cite{ondevicesurvey2021} and RSs \cite{scells2022reduce}. Common libraries like CodeCarbon \cite{schmidt2021codecarbon} lack support for edge devices. Additionally, our study is limited to Raspberry Pi and Python, which introduces considerable overhead and instability due to garbage collection. Future investigations could expand to include other programming languages and edge devices, enabling a more thorough benchmark of system efficiency. 
\end{itemize}

Additionally, we provide suggestions for practitioners on the effective application of LERS methods:
\begin{itemize}
\item \textbf{First}, in the CTR task, all compression techniques have drawbacks. The performance difference depends on dataset characteristics and sparsity. In our evaluation, pruning methods generally perform better on the Criteo dataset, while compositional embedding achieves better results on the Avazu dataset. Moreover, a method that performs well at one compression rate for a specific dataset is not guaranteed to perform well at another. However, in general, the difference between each method is minor. We therefore recommend starting with QR and PEP for the CTR task, as these methods have good training efficiency. Furthermore, in our tests, methods within the same category tend to demonstrate similar performance on a given dataset. After testing simpler methods, more complex ones within the best-performing category can be trained if necessary.

\item \textbf{Second}, in the CF task, graph-based models outperform latent-factor-based models; therefore, we suggest focusing on LERS graph-based models. Moreover, the performance drop is more significant in the CF task, likely due to the greater difficulty in compressing CF-based models. Nonetheless, every method still provides competitive results compared to the baseline and methods developed specifically for the CF task. This suggests that future research on LERSs should investigate performance in CF tasks in general and in graph-based models in particular.

\item \textbf{Third}, PEP typically outperforms other tested methods. Interestingly, simple approaches such as QR and MagPrune perform comparably to more complex ones, especially at lower sparsity rates. This suggests that practitioners should use simple methods when dealing with low sparsity rates, as these methods offer a good balance between simplicity and effectiveness.

\item \textbf{Fourth}, most methods are deployable on edge devices with reduced memory consumption. Both pruning and compositional approaches create runtime overhead, which varies based on the deployment environment and specific methods. In general, compositional methods appear to be a better choice for training efficiency, while pruning is more suitable for edge device settings, as CPUs can efficiently access unstructured weight matrices. Furthermore, a reduction in storage does not necessarily translate to lower memory costs in training. 
\end{itemize}


\begin{acks}
This work is supported by the Australian Research Council under the streams of Future Fellowship (No. FT210100624), Discovery Early Career Researcher Award (No. DE230101033), Discovery Project (No. DP240101108 and No. DP240101814), and Linkage Project (No. LP230200892).
\end{acks}

\appendix





\bibliographystyle{ACM-Reference-Format}
\bibliography{sample-base}

\end{document}